\documentclass{aa}  
\usepackage{graphicx}
\usepackage{txfonts}
\usepackage{multirow}
\usepackage{gensymb}
\usepackage{color}
\begin{document} 

\title{Exploring the climate of Proxima B with the Met Office Unified
  Model}


\author{Ian A. Boutle\inst{1,2}, 
Nathan J. Mayne\inst{2},
Benjamin Drummond\inst{2},
James Manners\inst{1,2}, 
Jayesh Goyal\inst{2},
F. Hugo Lambert\inst{3},
David M. Acreman\inst{2}
\and
Paul D. Earnshaw\inst{1}
}

\institute{Met Office, FitzRoy Road, Exeter, EX1 3PB, UK\\
\email{i.boutle@exeter.ac.uk}
\and
Physics and Astronomy, College of Engineering, Mathematics and
Physical Sciences, University of Exeter, Exeter, EX4 4QL, UK
\and
Mathematics, College of Engineering, Mathematics and
Physical Sciences, University of Exeter, Exeter, EX4 4QF, UK
}

\titlerunning{Climate of Proxima B}
\authorrunning{I. A. Boutle et al.}

\date{\today}

 
\abstract
{We present results of simulations of the climate of the newly
  discovered planet Proxima Centauri B, performed using the Met Office
  Unified Model (UM). We examine the responses of both an `Earth-like'
  atmosphere and simplified nitrogen and trace carbon dioxide
  atmosphere to the radiation likely received by Proxima Centauri
  B. Additionally, we explore the effects of orbital eccentricity on
  the planetary conditions using a range of eccentricities guided by
  the observational constraints. Overall, our results are in agreement
  with previous studies in suggesting Proxima Centauri B may well have
  surface temperatures conducive to the presence of liquid
  water. Moreover, we have expanded the parameter regime over which
  the planet may support liquid water to higher values of eccentricity
  ($\gtrsim 0.1$) and lower incident fluxes ($881.7$~W~m$^{-2}$)
  than previous work. This increased parameter space arises because of
  the low sensitivity of the planet to changes in stellar flux, a
  consequence of the stellar spectrum and orbital
  configuration. However, we also find interesting differences from
  previous simulations, such as cooler mean surface temperatures for
  the tidally-locked case. Finally, we have produced high resolution
  planetary emission and reflectance spectra, and highlight signatures
  of gases vital to the evolution of complex life on Earth (oxygen, ozone and
  carbon dioxide).}


\maketitle
%
\section{Introduction}
\label{section:introduction}

Motivated by the grander question of whether life on Earth is unique,
since the detection of the first exoplanet \cite[]{WolF92,MayQ95},
efforts have become increasingly focused on `habitable' planets. In
order to partly side-step our ignorance of the possibly vast range of
biological solutions to life, and exploit our mature understanding of
our own planet's climate, we define the term `habitable' in a very
Earth-centric way. Informed by the `habitable zone' first defined by
\cite{Kas88}, we have searched for planets where liquid water, so
fundamental to Earth-life, might be present on the planetary
surface. The presence of liquid water is likely to depend on a large
number of parameters such as initial water budget, atmospheric
composition, flux received from the host star etc. Of course, the more
`Earth-like' the exoplanet, the more confidence we have in applying
adaptations of our theories and models developed to study Earth
itself, and thereby predicting parameters such as surface temperature and
precipitation. Surveys such as the Terra Hunting Experiment
\cite[]{ThoQB16} aim to discover targets similar in both mass and
orbital configuration to Earth, but also orbiting stars similar to our
Sun, for which follow-up characterisation measurements might be
possible.

In the meantime, observational limitations have driven the search for
`Earth-like' planets to lower mass, and smaller radius stars
i.e. M-Dwarfs \cite[e.g. MEarth,][]{NutC08}. Such stars are much
cooler and fainter than our own Sun, so potentially habitable planets
must exist in much tighter orbits. Recent, ground breaking, detections
have been made for potentially ``Earth-like'' planets in orbit around
M-Dwarfs { (e.g.~Gliese 581g \cite[]{VogtBR10}, Kepler 186f \cite[]{QuintanaBR14}, the Trappist 1 system \cite[]{GilJL16})}. In fact, such
a planet has been discovered orbiting, potentially in the `habitable
zone' of our nearest neighbour Proxima Centauri \cite[]{AngABB16},
called Proxima Centauri B (hereafter, ProC B).

The announcement of this discovery was coordinated with a
comprehensive modelling effort, exploring the possible effects of the
stellar activity on the planet over its evolution and its budget of
volatile species \cite[]{RibBS16}, and a full global circulation model
(GCM) of the climate \cite[]{TurLS16}. Of course, lessons from solar
system planets { \cite[]{ForL13}} and our own Earth-climate
\cite[]{FlaM13} have taught us that the complexity of GCMs can lead to
model dependency in the results. This can often be due to subtle
differences in the numerics, various schemes (i.e. radiative transfer,
chemistry, clouds etc.) or boundary and initial
conditions. \cite{Gol16} provides an excellent resource, using 1D
models and limiting concepts with which to aid conceptual
understanding of the habitability of ProC B, highly complementary to
results from more complex 3D models such as \cite{TurLS16} and this
work.

In this work we apply a GCM of commensurate pedigree and
sophistication to that used by \cite{TurLS16} to ProC B and explore
differences due to the model, and extend the exploration to a wider
observationally plausible parameter space. The GCM used is the Met
Office Unified Model (or UM), which has been successfully used to
study Earth's climate for decades. We have adapted this model,
introducing flexibility to enable us to model a wider range of
planets. Our efforts have focused on gas giant planets
\cite[]{MayBA14aa,AmuMB16}, motivated by observational constraints,
but have included simple terrestrial Earth-like planets
\cite[]{MayBA14gmd}.

The structure of the paper is as follows: in Section
\ref{section:model_setup} we detail the model used, and the parameters
adopted. For this work we focus on two cases, an Earth-like atmosphere
chosen to explore how an idealised Earth climate would behave under
the irradiation conditions of ProC B, and a simple atmosphere
consisting of nitrogen with trace CO$_2$ for a cleaner comparison with
the work of \cite{TurLS16}. In Section \ref{section:results} we
discuss the output from our simulations, and compare them to the
results of \cite{TurLS16}, revealing a slightly cooler day-side of the
planet in the tidally-locked case (likely driven by differences in the
treatment of clouds, convection, boundary layer mixing, and vertical
resolution), and a warmer mean surface temperature for the 3:2
spin-orbit resonance configuration, { particularly} when adopting an eccentricity of
$0.3$ \cite[compared to zero in][]{TurLS16}. Our simulations suggest
that the mean surface temperatures move above the freezing point of
water for eccentricities of around { 0.1} and greater. Section
\ref{section:spectra} presents reflection (shortwave) spectra,
emission (longwave) spectra and reflection/emission as a function of
{ time and orbital phase angle} derived from our simulations. Our results show
many similar trends to the results of \citet{TurLS16}, with several
important differences. In particular, our model is capable of a higher
spectral resolution, allowing us to highlight the spectral signature
of the gases key to the evolution of complex life on
Earth (ozone, oxygen, carbon dioxide).

Finally, in Section \ref{section:conclusions} we conclude that the
agreement between our simulations and those of \cite{TurLS16} further
confirms the \textit{potential} for ProC B to be
\textit{habitable}. However, the discrepancies mean further
inter-comparison of detailed models is required, and must always be
combined with the insight provided by 1D, simplified approaches such
as \cite{Gol16}.

\section{Model Setup}
\label{section:model_setup}

The basis of the model simulations presented here is the Global
Atmosphere (GA) 7.0 \cite[]{WalBB17} configuration of the Met Office
Unified Model. This configuration will form the basis of the Met
Office contribution to the next Intergovernmental Panel on Climate
Change (IPCC) report, and will replace the current GA6.0
\cite[]{WalBBM17} configuration for operational numerical weather
prediction in 2017. It therefore represents one of the most
sophisticated and accurate models for Earth's atmosphere, and with
minimal changes can be adapted for any Earth-like atmosphere. The
model solves the full, deep-atmosphere, non-hydrostatic, Navier-Stokes
equations using a semi-implicit, semi-Lagrangian approach. It contains
a full suite of physical parametrizations to model sub-grid scale
turbulence (including non-local turbulent transport), convection
(based on a mass-flux approach), H$_2$O cloud and precipitation
formation (with separate prognostic treatment of ice and liquid
phases) and radiative transfer. Full details of the model dynamics and
physics can be found in \cite{WalBBM17} and \cite{WalBB17}. The
simulations presented have a horizontal resolution of $2.5\degree$
longitude by $2\degree$ latitude, with $38$ vertical levels between
the surface and model-top (at $40$~km), quadratically stretched to
give enhanced resolution near the surface. We adopt a timestep of
$1200$~s.

\begin{table}[tbhp]
  \caption[]{Orbital and planetary parameters used in this study, for
    spin-orbit resonances of 1:1 and 3:2.}
  \label{tab-planet}
      $$
      \begin{array}{p{0.48\linewidth}|p{0.205\linewidth}|p{0.205\linewidth}}
        Parameter & 1:1 & 3:2 \\
        \hline
        Semi-major axis / AU & \multicolumn{2}{c}{0.0485} \\
        Stellar irradiance / W~m$^{-2}$ (S$_\oplus$) & \multicolumn{2}{c}{881.7~(0.646)}\\
        Orbital period / Earth days & \multicolumn{2}{c}{11.186}\\
        $\Omega$ / rad~s$^{-1}$ & $6.501\cdot10^{-6}$ & $9.7517\cdot10^{-6}$\\
        Eccentricity & $0$ & $0.3$\\
        Obliquity & \multicolumn{2}{c}{0} \\
        Radius / km (R$_\oplus$) & \multicolumn{2}{c}{7160~(1.1)}\\
        $g$ / m~s$^{-2}$ & \multicolumn{2}{c}{10.9}\\
      \end{array}
      $$
\end{table}

To adapt the model for simulations of ProC B, we modify the planetary
parameters to those listed in Table~\ref{tab-planet}. The orbital
parameters are taken from \cite{AngABB16} and we note that our values
for the stellar irradiance and rotation rate ($\Omega$) differ to
those used by \cite{TurLS16}. In particular, our value for the stellar
irradiance ($881.7$~W~m$^{-2}$), based on the best estimates for the
stellar flux of Proxima Centauri and semi-major axis of ProC B, is
considerably lower than the $956$~W~m$^{-2}$ used in
\cite{TurLS16}. The planetary parameters (radius and $g$) are taken
from \cite{TurLS16}. 

\begin{figure}[tbhp]
  \centering
  \includegraphics[width=\columnwidth]{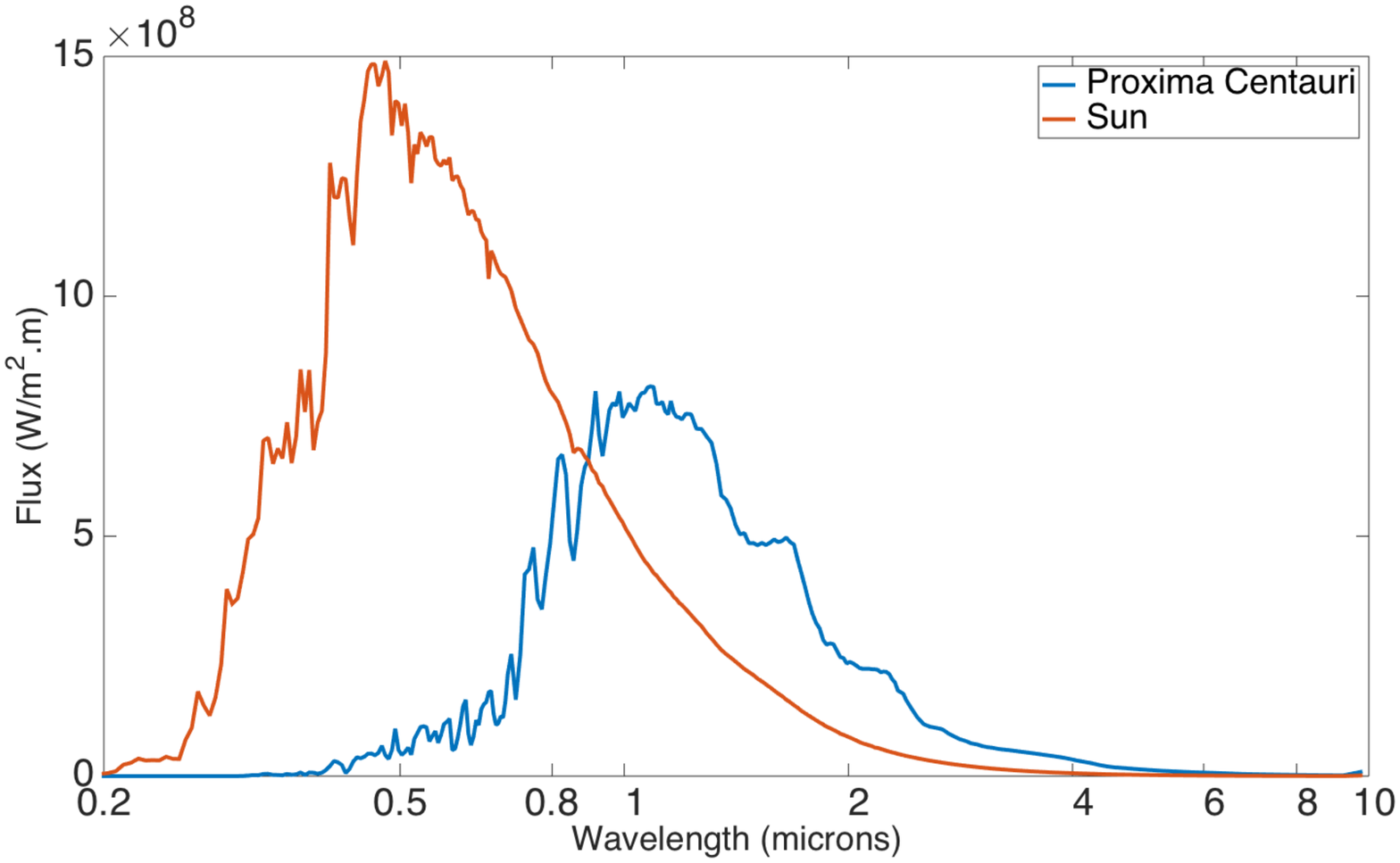}
  \caption{Top of atmosphere spectral flux for Earth (red) and ProC B
    (blue).}
  \label{fig-spectra}
\end{figure}

The SOCRATES\footnote{https://code.metoffice.gov.uk/trac/socrates}
radiative transfer scheme is used with a configuration based on the
Earth's atmosphere \citep[GA7.0,][]{WalBB17}. Incoming stellar
radiation is treated in 6 "shortwave" bands ($0.2-10$~$\mu$m), and
thermal emission from the planet in 9 "longwave" bands ($3.3$~$\mu$m
-- $10$~mm), applying a correlated-$k$ technique. Absorption by water
vapour and trace gases, Rayleigh scattering, and absorption and
scattering by liquid and ice clouds are included. The clouds
themselves are water-based, and modelled using the PC2 scheme which is
detailed in \citet{WilBKP08}. Adaptations are made to represent the
particular stellar spectrum of Proxima Centauri. A comparison of the
top of atmosphere spectral flux for Earth and ProC B is shown in
Figure~\ref{fig-spectra}.
The Proxima Centauri stellar spectrum is from BT-Settl
\cite[]{RajRA13} with $T_{\rm eff}=3000$~K, $g=1000$~m~s$^{-2}$ and
metallicity$=0.3$~dex, based on \cite{SchL10}. Correlated-$k$
absorption coefficients are recalculated using this stellar spectrum
to weight wavelengths within the shortwave bands, and to cover the
wider range of temperatures expected on ProC B. For
simplicity we ignore the effects of atmospheric aerosols in all
simulations, although tests with a simple representation of aerosol
absorption and scattering did not lead to a significant difference in
results. For the spectra and phase curves presented in
Section~\ref{section:spectra} we additionally run short GCM
simulations with high resolution spectral files containing 260
shortwave and 300 longwave bands that have been similarly adapted
for ProC B from the original GA7 reference configurations.

Similar to \cite{TurLS16}, we use a flat, homogeneous surface at our
inner boundary, but for simplicity choose a single layer `slab' model
based on \cite{FriHZ06}. The heat-capacity of
$10^7$~J~K$^{-1}$~m$^{-2}$ is representative of a sea-surface with
$2.4$~m mixed layer, although as all simulations are run to
equilibrium, this choice does not affect the mean temperature, only
the variability (capturing the diurnal cycle). We consider that
simulations have reached equilibrium when the top-of-atmosphere is in
radiative balance, the hydrological cycle (surface precipitation minus
evaporation) is in balance and the stratospheric temperature is no
longer evolving. We find that equilibrium is typically reached within
30 orbits, and show most diagnostics averaged over orbits 80-90
(sampled every model timestep). We retain water-like properties of the
surface (even below $0\degree$C) allowing the roughness length to vary
with windspeed, typically between $10^{-5}$ and $10^{-3}$~m. The
emissivity of the surface is fixed at $0.985$ and the albedo varies
with stellar zenith angle, ranging from $0.05$ at low zenith angles
but reaching $0.5$ at very high zenith angles.

\begin{table}[tbhp]
  \caption[]{Atmospheric parameters used in this study, for
    nitrogen dominated and Earth-like atmospheric compositions, where the Earth-like case contains further trace gases as in Earth's current atmosphere.}
  \label{tab-atm}
     $$
     \begin{array}{p{0.32\linewidth}|p{0.32\linewidth}|p{0.25\linewidth}}
       Parameter & nitrogen+trace CO$_2$ & Earth-like\\
       \hline
       $R$ / J~kg$^{-1}$~K$^{-1}$ & $297$ & $287.05$ \\
       $c_p$ / J~kg$^{-1}$~K$^{-1}$ & $1039$ & $1005$ \\
       CO$_2$ MMR / kg~kg$^{-1}$ & \multicolumn{2}{c}{5.941\cdot10^{-4}} \\
       O$_2$ MMR / kg~kg$^{-1}$ & $0$ & $0.2314$ \\
       \multirow{2}{*}{O$_3$ MMR / kg~kg$^{-1}$} & \multirow{2}{*}{$0$} & $2.4\cdot10^{-8}$ (min)\\
       & & $1.6\cdot10^{-5}$ (max)\\
       CH$_4$ MMR / kg~kg$^{-1}$ & $0$ & $1.0\cdot10^{-7}$\\
       N$_2$O MMR / kg~kg$^{-1}$ & $0$ & $4.9\cdot10^{-7}$\\
      \end{array}
      $$
\end{table}

All simulations have an atmosphere with a mean surface pressure of
$10^{5}$~Pa, and we investigate two different atmospheric
compositions, the relevant parameters for which are given in
Table~\ref{tab-atm}. These represent a nitrogen dominated atmosphere
with trace amounts of CO$_2$, similar to that investigated by
\cite{TurLS16}, and a more Earth-like atmospheric composition with
significant oxygen and trace amounts of other radiatively important
gases. Our motivation here is to explore the possible climate and
observable differences that would exist on a planet that did support
complex life \cite[]{LenW11}. The values for the gases are taken from present
day Earth, and are globally uniform with the exception of ozone, for
which we apply an Earth-like distribution, with highest values in the
equatorial stratosphere, decreasing towards the poles and with much
lower values in the troposphere. Whether an ozone layer could form and
survive on ProC B is highly uncertain. Ozone formation requires
radiation with wavelengths of $0.16-0.24\mu$m, which we expect to be
in much shorter supply for ProC B, compared with Earth; see
Figure~\ref{fig-spectra} and \cite{MeaAS18}. \cite{MeaAS18} also
discuss that the likelihood of stellar flares destroying the ozone
layer is quite high, and without it the chances of habitability are
significantly reduced due to the large stellar fluxes at very short
wavelengths ($<0.175\mu$m) received by ProC B. Essentially, in this
work, our main aim is to investigate the response of an ``Earth-like''
atmosphere to the irradiation conditions (different spectrum and
stellar flux patterns) characteristic of the ProC B system, so we
refrain from removing individual gases which may actually be required
for the planet to be habitable, or are potentially produced by an
interaction of life with the atmosphere, such as ozone and methane
\cite[]{LenW11}. A 3D model fully-consistent with the chemical
composition is beyond the scope of the present work.

%

\section{Results}
\label{section:results}

In this section we discuss results from our simulations in two orbital
configurations. Firstly, the assumption of a tidally-locked planet,
and then a 3:2 spin-orbit resonance, both possible for such a planet
as ProC B { \cite[]{RibBS16}}.

\subsection{Tidally-locked case}
\label{sec-fixed}

\begin{figure}[tbhp]
  \centering
  \includegraphics[width=\columnwidth]{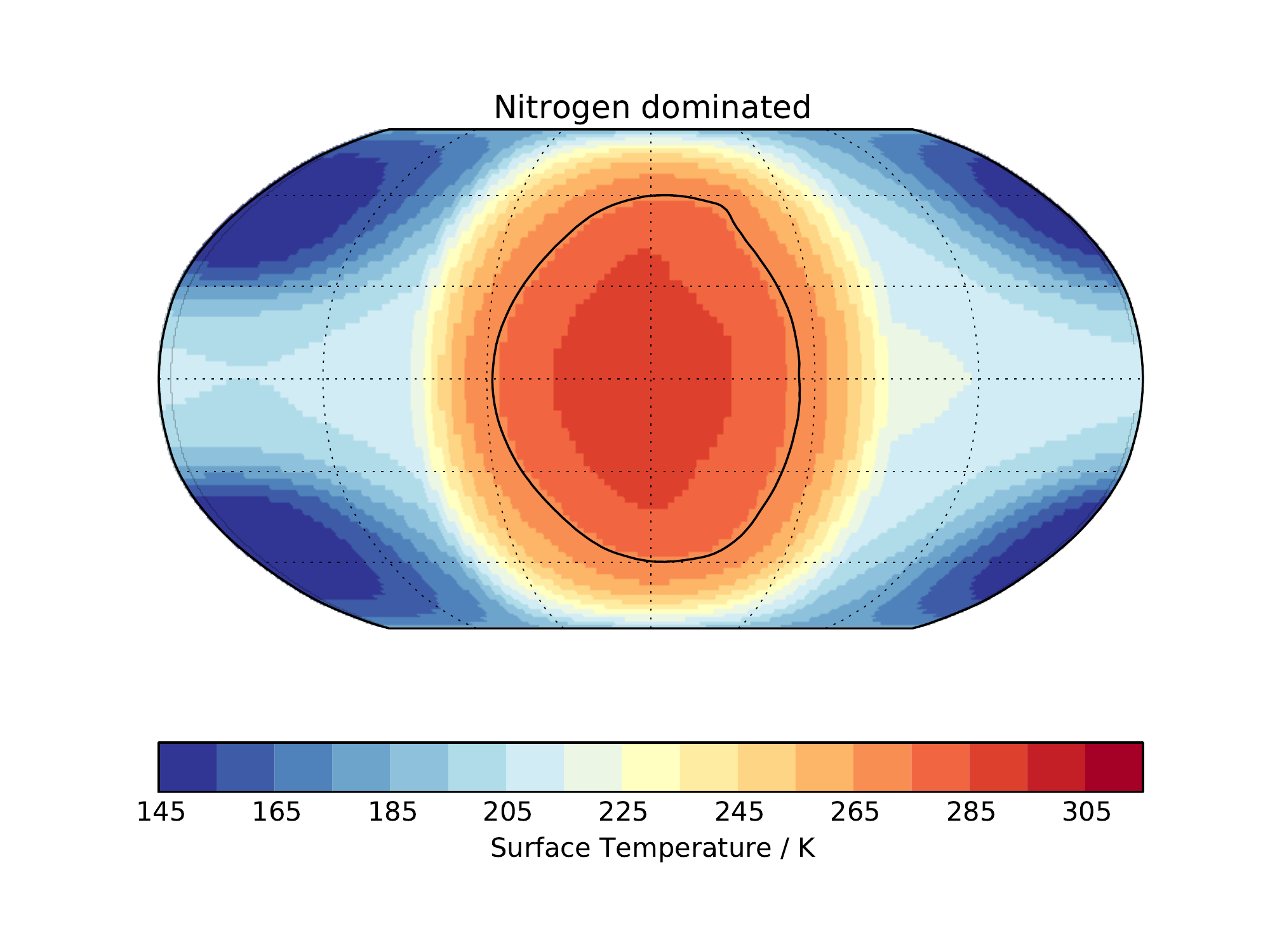}
  \includegraphics[width=\columnwidth]{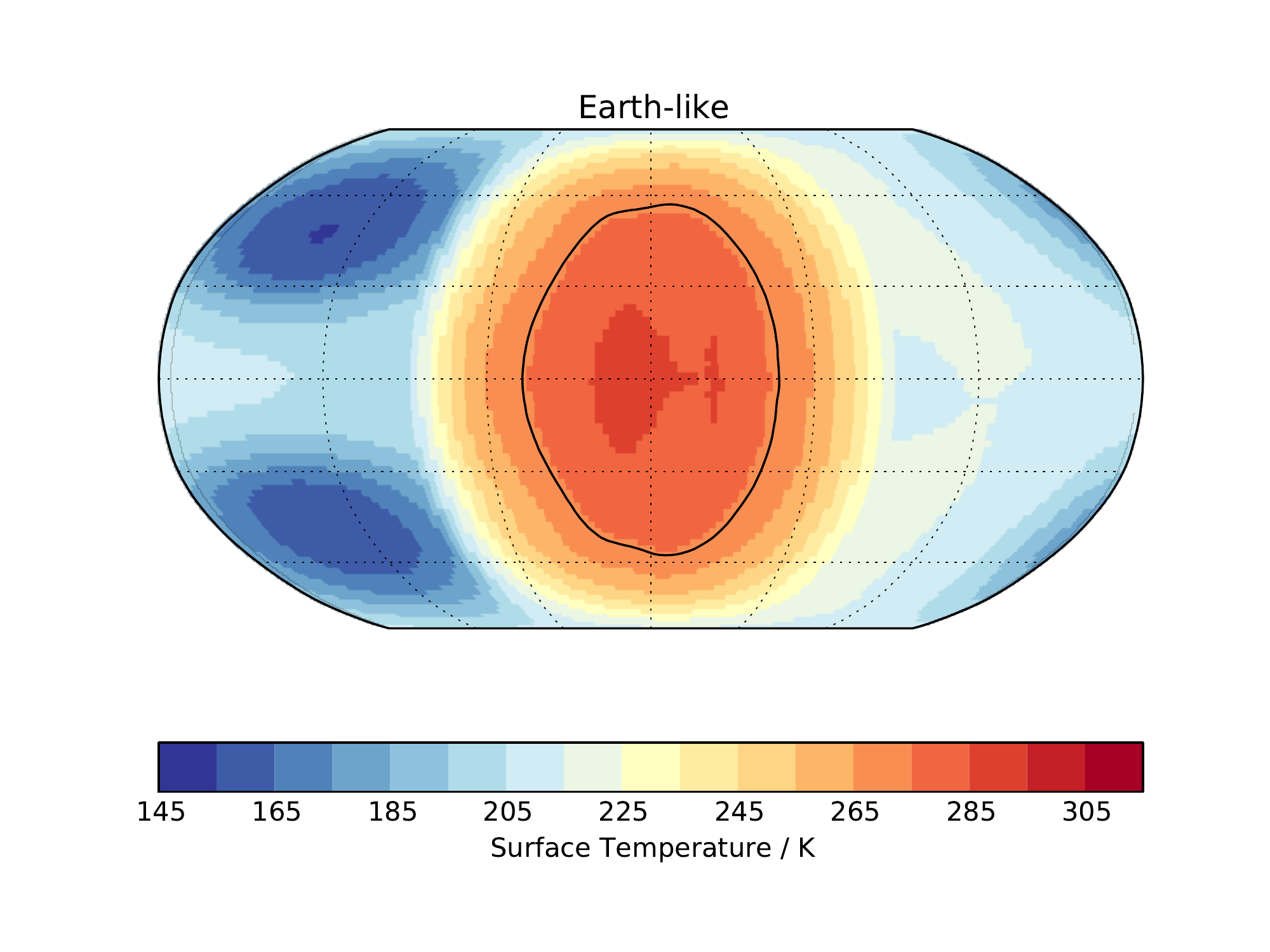}
  \caption{{ 10 orbit mean} surface temperature (coloured) with the mean $0\degree$C
    contour, for the tidally-locked nitrogen dominated and Earth-like
    atmospheric compositions.}
  \label{fig-surft}
\end{figure}

We first consider the tidally-locked orbit with zero
eccentricity. Figure~\ref{fig-surft} shows the surface temperature
from our simulations. It is colder than the simulations of
\cite{TurLS16}, with a maximum temperature on the day-side of $290$~K
($10$~K colder), and a minimum temperature in the cold-traps on the
night-side of { $150$~K, ($50$~K colder, informed by their
figures)}. There are several reasons for these differences which we
will explore.

Firstly, we adopt a stellar radiation at the top of the atmosphere
which is $70$~W~m$^{-2}$ lower than \cite{TurLS16}, as discussed in
Section~\ref{section:model_setup}, and so will inevitably be
colder. We have tested our model with an incoming stellar flux
consistent with that used by \cite{TurLS16}, and find that it
increases the mean surface temperature by $5$~K across the
planet { (slightly less on the day-side and up to $10$~K in the cold-traps on the night side)} which, critically, is still cooler than that found by
\cite{TurLS16}. This increase is approximately two-thirds of that
found for Earth \cite[$4$~K for $45$~W~m$^{-2}$ additional solar
flux,][]{AndRD12}, demonstrating that the sensitivity of planetary
temperatures to changes in the stellar flux received by ProC B is quite
low, meaning it potentially remains habitable over a larger range of orbital radii than e.g.~Earth. This is
likely to be due to a combination of the tidal locking and stellar
spectrum. { For example, changes in low cloud and ice amounts that contribute to a strong shortwave feedback on Earth are ineffective in this configuration as low clouds and ice are found largely on the night side of the planet}.

\begin{figure}[tbhp]
  \centering
  \includegraphics[width=\columnwidth]{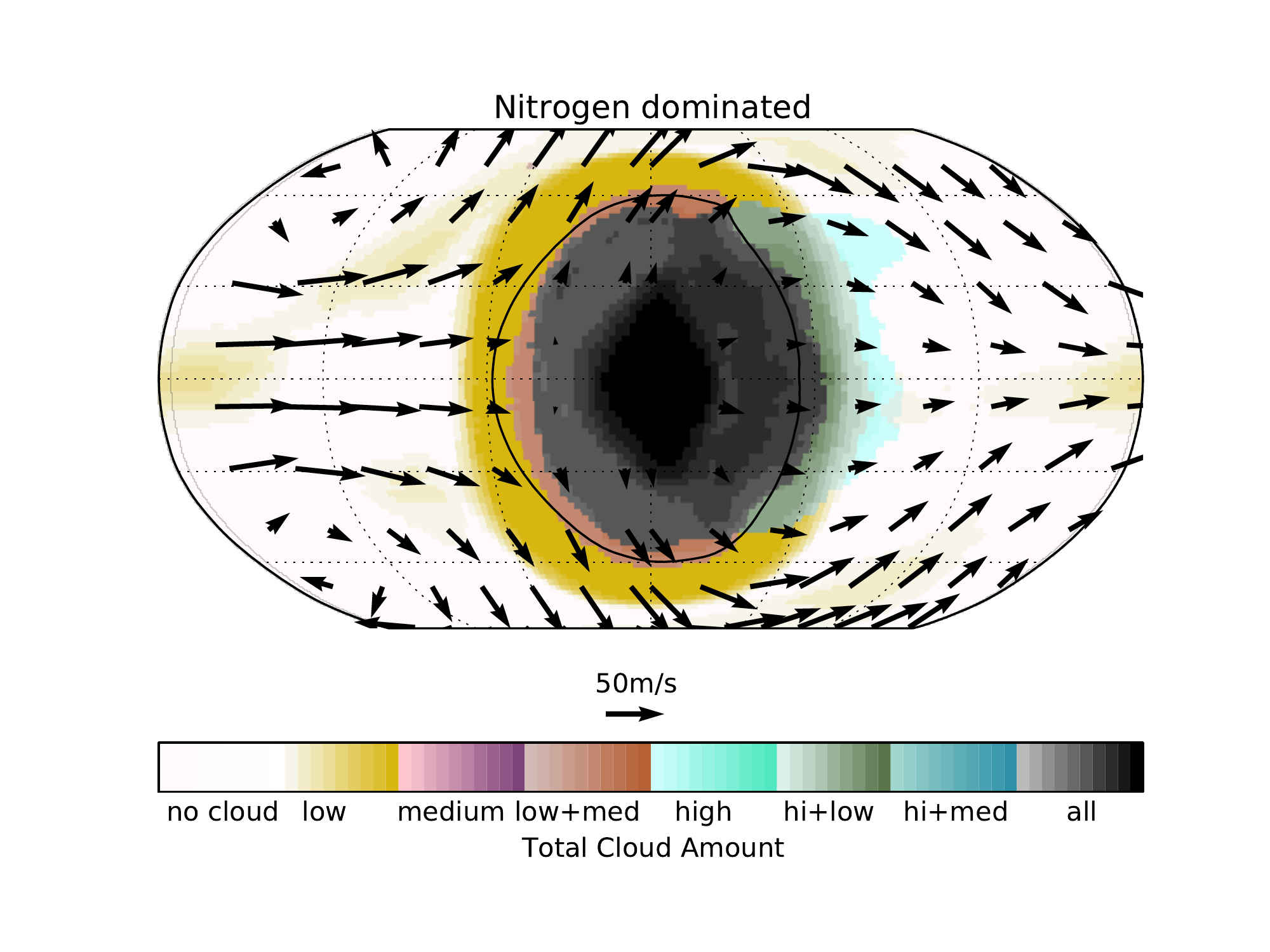}
  \includegraphics[width=\columnwidth]{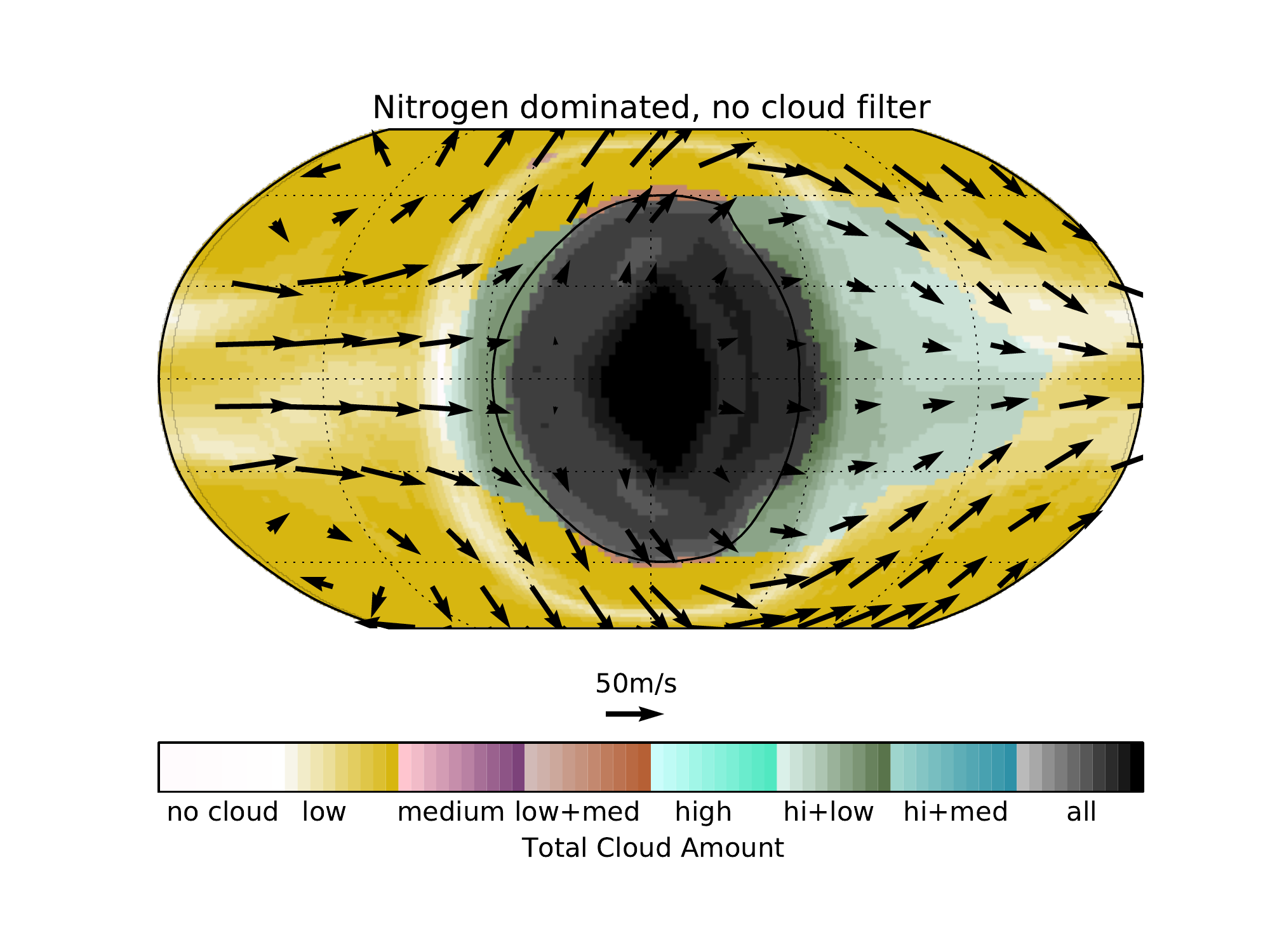}
  \caption{{ 10 orbit mean} cloud cover (coloured) with the mean $0\degree$C surface
    temperature contour and wind-vectors at $8.5$~km, for the nitrogen
    dominated tidally-locked simulations, with and without filtering
    of optically thin cloud. { Within each altitude range (low < $2$~km <  medium < $5.5$~km < high) the cloud cover is given by the maximum cloud fraction ($\in [0,1]$) on any model level. If only one type of cloud is present, the colourbar shows that fraction, with a contour interval of $0.1$. If more than one type of cloud is present, the colourbar shows the average of the two or three cloud types present, again with colourbar interval of $0.1$.}}
  \label{fig-cloud}
\end{figure}

On the day-side, cloud cover could be a contributing factor in keeping
the surface cooler in our simulations. As shown in
Figure~\ref{fig-cloud}, the day-side of the planet is completely
covered in cloud, due to the strong stellar heating driving convection
and cloud formation. This makes the albedo of the day-side quite high
($\approx 0.35$), reflecting a significant fraction of the incoming
radiation back to space, similar to simulations presented by
\cite{YanCA13}.
\begin{figure}[tbhp]
  \centering
  \includegraphics[width=\columnwidth]{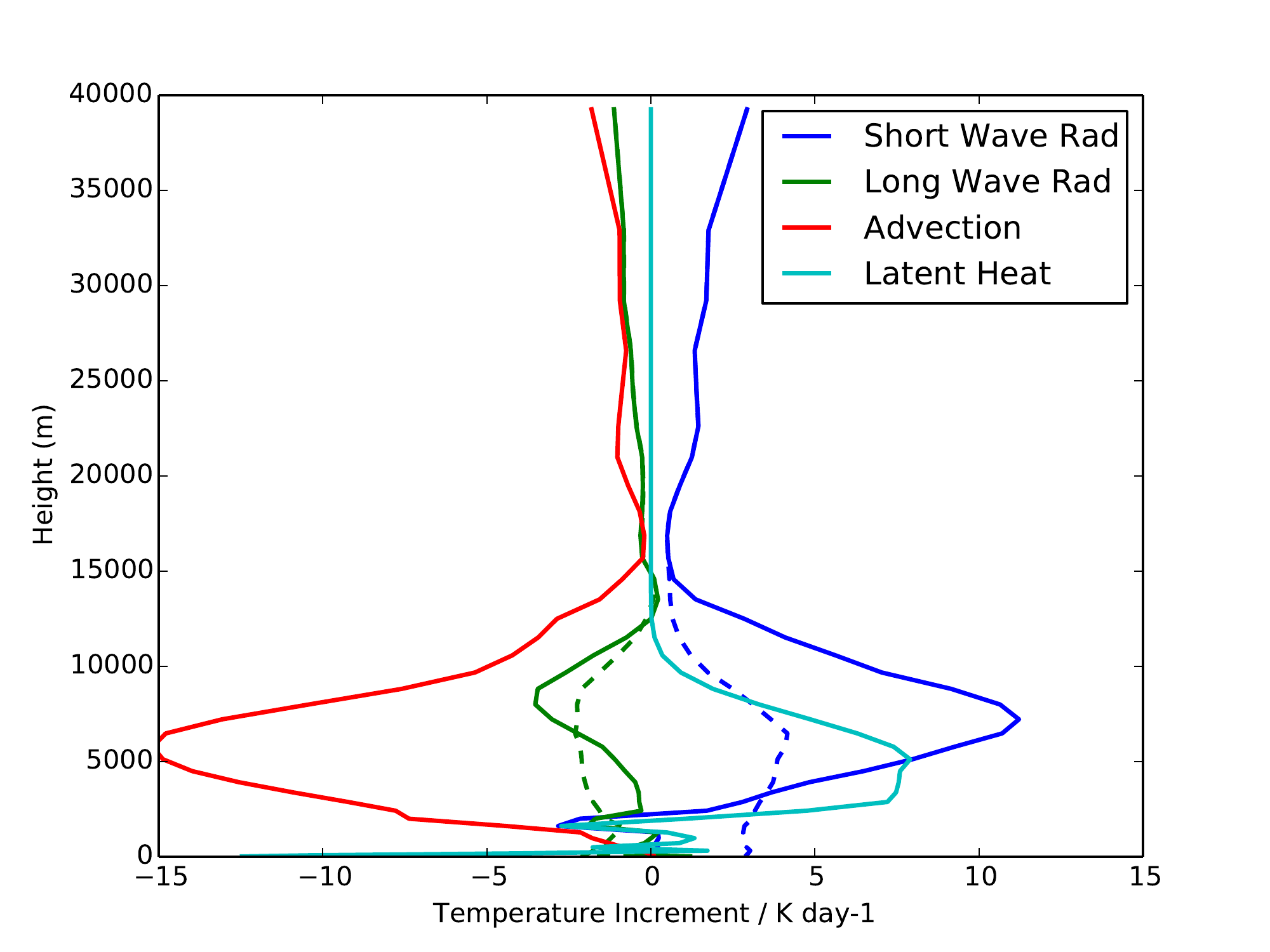}
  \caption{{ 10 orbit mean} heating profiles from the sub-stellar point on the day-side
    of the nitrogen dominated tidally-locked simulation for the main
    physical processes. The dashed lines show the clear-sky heating
    from the radiation components.}
  \label{fig-heat}
\end{figure}
Furthermore, the radiative heating of the thick cloud layer which
forms is very high ($>10$~K~day$^{-1}$, Fig.~\ref{fig-heat}). It is
possible that the combination of these two effects is greater in our
model (driven by differences in our cloud and convection schemes,
discussed later in this section), simply resulting in less radiation
reaching the planet surface, and therefore a cooler surface
temperature. However, the cooler day-side temperature may actually be
linked to the temperature on the night-side, via the mechanisms
described in \cite{YanA14}. They argue that the free tropospheric
temperature should be horizontally uniform, due to the global-scale
Walker circulation that exists on a tidally-locked planet
\cite[]{ShoWM13}, and efficient redistribution of heat by the
equatorial superrotating jet \cite[]{ShoP11}. Figure~\ref{fig-prof}
shows this to be true in our simulations, and the weak temperature
gradient \cite[]{Pie95} effectively implies that the temperature of
the entire planet is controlled by the efficiency with which emission
of longwave radiation to space can cool the night-side of the
planet. Therefore, the fact that our night-side is so cold implies a
very efficient night-side cooling mechanism which in turn suppresses
the day-side temperatures.

The temperature on the night-side is cold due to the almost complete
absence of cloud and very little water vapour. This allows the surface
to continually radiate heat back to space, and cool
dramatically. The only mechanism to balance this heat loss is
transport from the day-side of the planet at higher levels within the
atmosphere, followed by subsidence (where a layer of air descends and heats
under compression) or sub-grid mixing to transport the heat down to
the surface. Figure~\ref{fig-prof} shows profiles of temperature from
the day- and night-side of the planet, demonstrating that the cooling
is confined to the lowest { $3$~km} of the atmosphere, with the most
extreme cooling { ($30$~K)} in the lowest $1$~km. We speculate that it is
this near surface cooling which differs between our model and that of
\cite{TurLS16}, as our temperature at $500$~m altitude appears very
similar to their surface temperature (not shown).

\begin{figure}[tbhp]
  \centering
  \includegraphics[width=\columnwidth]{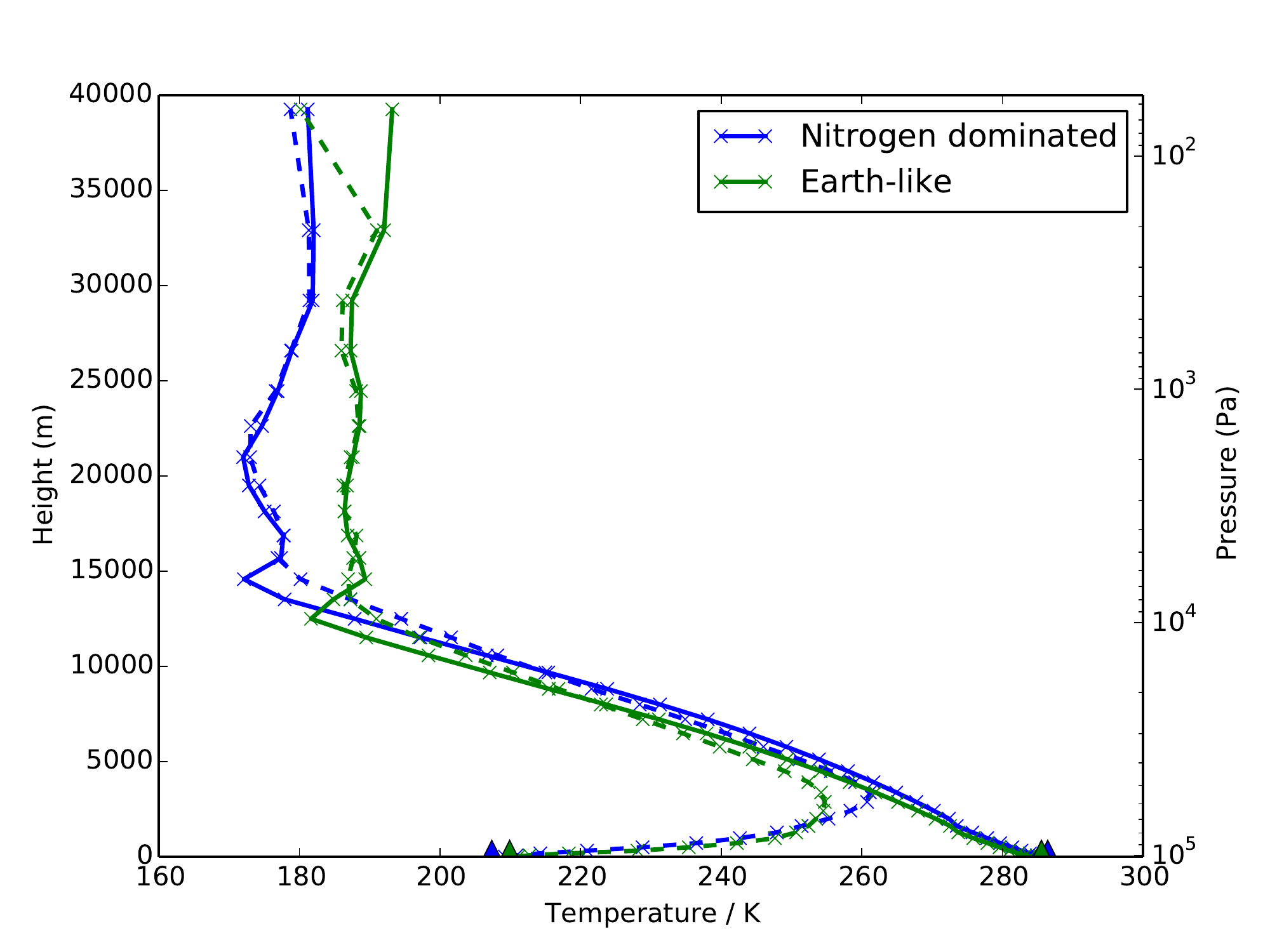}
  \includegraphics[width=\columnwidth]{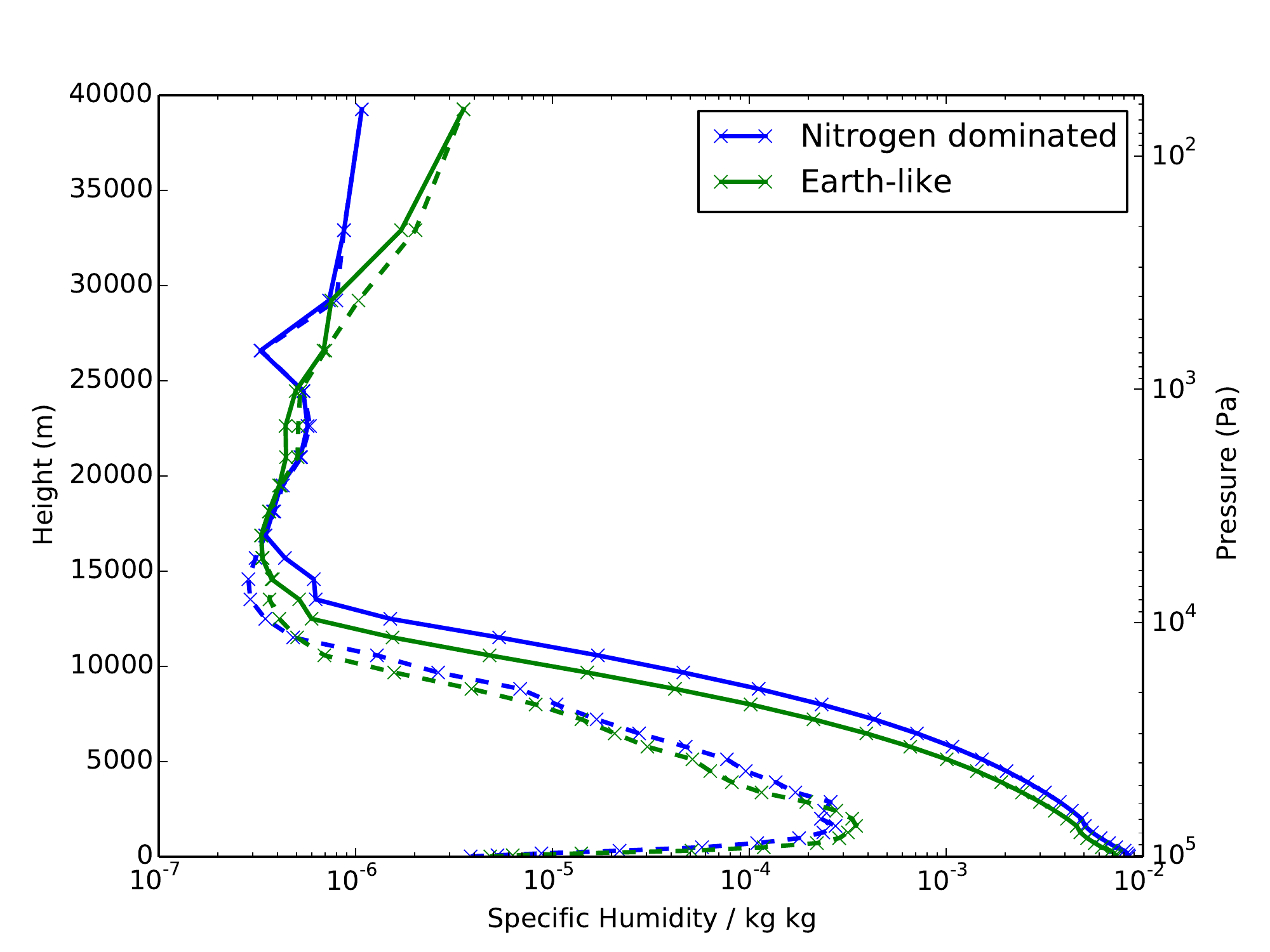}
  \caption{{ 10 orbit mean} profiles of temperature and water vapour (\textit{top} and
    \textit{bottom panels}, respectively), from the sub-stellar point
    on the day-side of the tidally-locked planet (solid), and its
    antipode (dashed). Filled triangles show the surface temperature,
    and crosses indicate the position of the model-levels.}
  \label{fig-prof}
\end{figure}

There are several possible reasons for the surface temperature
differences between our model and that of \cite{TurLS16}. Firstly, the
water-vapour profile could play a role. The night-side is dry because
its only real source of water vapour is transport from the day-side,
but this transport typically happens at high levels within the
atmosphere, where the air is very dry due to the efficiency with which
the deep convection precipitates water. This is likely to be a key
uncertainty and potential reason for differences between simulations,
as the convective parametrizations are very different -- a simple
adjustment scheme \cite[]{ManW67} used in \cite{TurLS16} versus a
mass-flux based transport scheme \cite[based on][but with significant improvements]{GreR90} used here. Secondly,
model-resolution and the parametrization of turbulent mixing in the
stable atmosphere are hugely important. How much sub-grid mixing
atmospheric models should apply in stable regions is still a topic of
research in the GCM community \cite[]{HolSB13}, with many GCMs often
applying more mixing than observations or theory would suggest. The UM
uses a minimal amount of mixing in stable regions, which results in
very little transport of heat down to the surface by sub-grid
processes, and relies on the subsidence resolved on the model grid to
warm the surface, which is also very weak in our lowest model level
($20$~m above the surface). Tests with increased mixing can produce a
$20$~K increase in surface temperature, and also significantly alter
the positions of the cold-traps.

The absence of cloud is another possible reason for surface
temperature differences; results presented in \cite{YanCA13} showed
uniform low level cloud cover on the night-side of a tidally-locked
planet, which could help to insulate the surface and keep it
warm. However, what cloud there is on the night-side of our model has
such low water content that it is optically very thin and has almost
no effect on the radiation budget. In Figure~\ref{fig-cloud}, we show
the same cloud cover field, but in the bottom panel we show cloud as
any grid-box with condensed water, whereas in the top panel we only
consider a grid-box to be cloudy if that cloud is radiatively
important (e.g. it would be visible to the human eye). This is done by
filtering all cloud with an optical depth $<0.01$ from the
diagnostic. This shows that whilst the cloud cover can appear quite
extensive on the night-side, the cloud is actually radiatively
unimportant. Finally, our model is lacking a representation of
condensible CO$_2$, which could be an important contributor to the
radiative balance of the night-side, both if vapour CO$_2$
concentrations are locally increased on the night-side, or CO$_2$
clouds are present. { However, for the concentrations of CO$_2$
  considered here, condensation would occur
  at $\approx 125$~K near the surface, and therefore condensation of
  CO$_2$ would appear unlikely, even in the cold-traps}. We note that our surface temperature on the
night-side appears very similar to the dry case of \cite{TurLS16}, and
that our night-side surface temperature appears to match the very cold
results given by the simple model of \cite{YanA14} better than their
GCM results which kept the surface warmer.

The temperature and water-vapour profiles shown in
Figure~\ref{fig-prof} appear in good agreement with \cite{TurLS16}.
{ Figure~\ref{fig-heat} shows that there is significant shortwave heating in the stratosphere, a result of}
shortwave absorption by CO$_2$ in our model, which is happening
longward of $2\mu$m. This is a feature of Proxima Centauri's spectrum
(Fig~\ref{fig-spectra}), and would not happen on solar system planets
due to the much lower flux at this wavelength from the
Sun. The
heating is balanced by longwave cooling from the CO$_2$ and water
vapour, and transport of heat to the night-side of the
planet. Figure~\ref{fig-heat} shows that heat transport is the
dominant mechanism of heat-loss from the day-side throughout the
atmosphere, and this heat is transported to the night-side where it is
the only heat source and balanced by longwave cooling.

The differences due to atmospheric composition are generally quite
small within the troposphere. The Earth-like composition has a similar
surface temperature on the day-side, and slightly warmer surface
temperature on the night-side, particularly in the cold
traps. Consistent with \cite{YanA14}, this difference is primarily
driven by additional heat on the day-side of the planet being
transported to the night-side, effectively stabilising the temperature
of the day-side and increasing the temperature of the night-side. Most
other fields are very similar and not shown for brevity. There are
however significant differences in the stratosphere
(Fig.~\ref{fig-prof}). The stratosphere is { warmer}, and this is
predominantly driven by the ozone layer. { However, the warming created is much less than on Earth, because there is very little}
radiative flux in the region which ozone absorbs
($0.2-0.32\mu$m). The
stratosphere is also wetter, and this is a direct consequence of water
vapour production by methane oxidation in this configuration. { This is achieved via a simple parametrization \cite[]{UntS98}, common in many GCMs, that increases stratospheric water vapour in proportion to the assumed methane mixing ratio and observed balance between water vapour and methane in Earth's stratosphere.}

\begin{figure}[tbhp]
  \centering
  \includegraphics[width=\columnwidth]{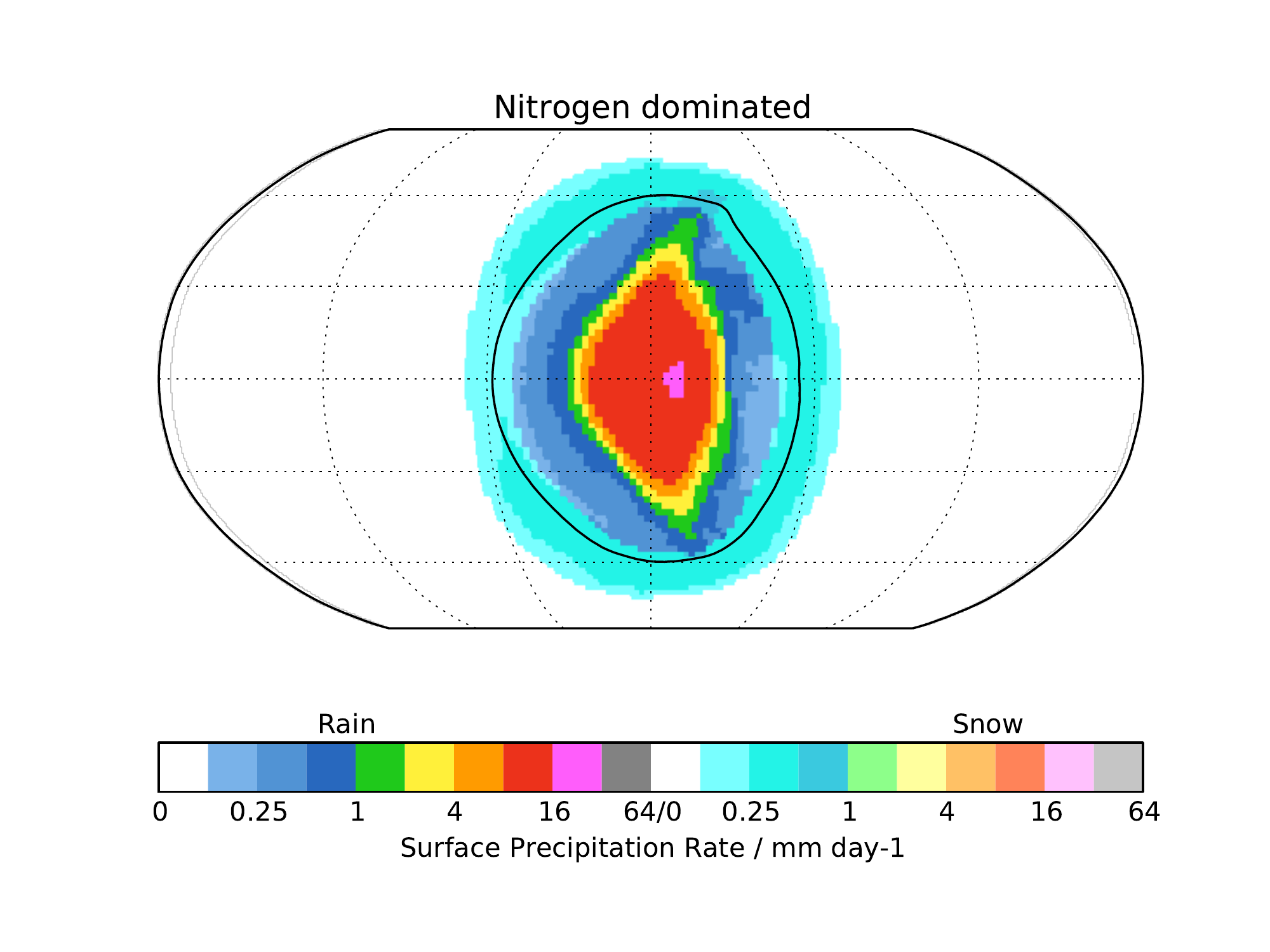}
  \includegraphics[width=\columnwidth]{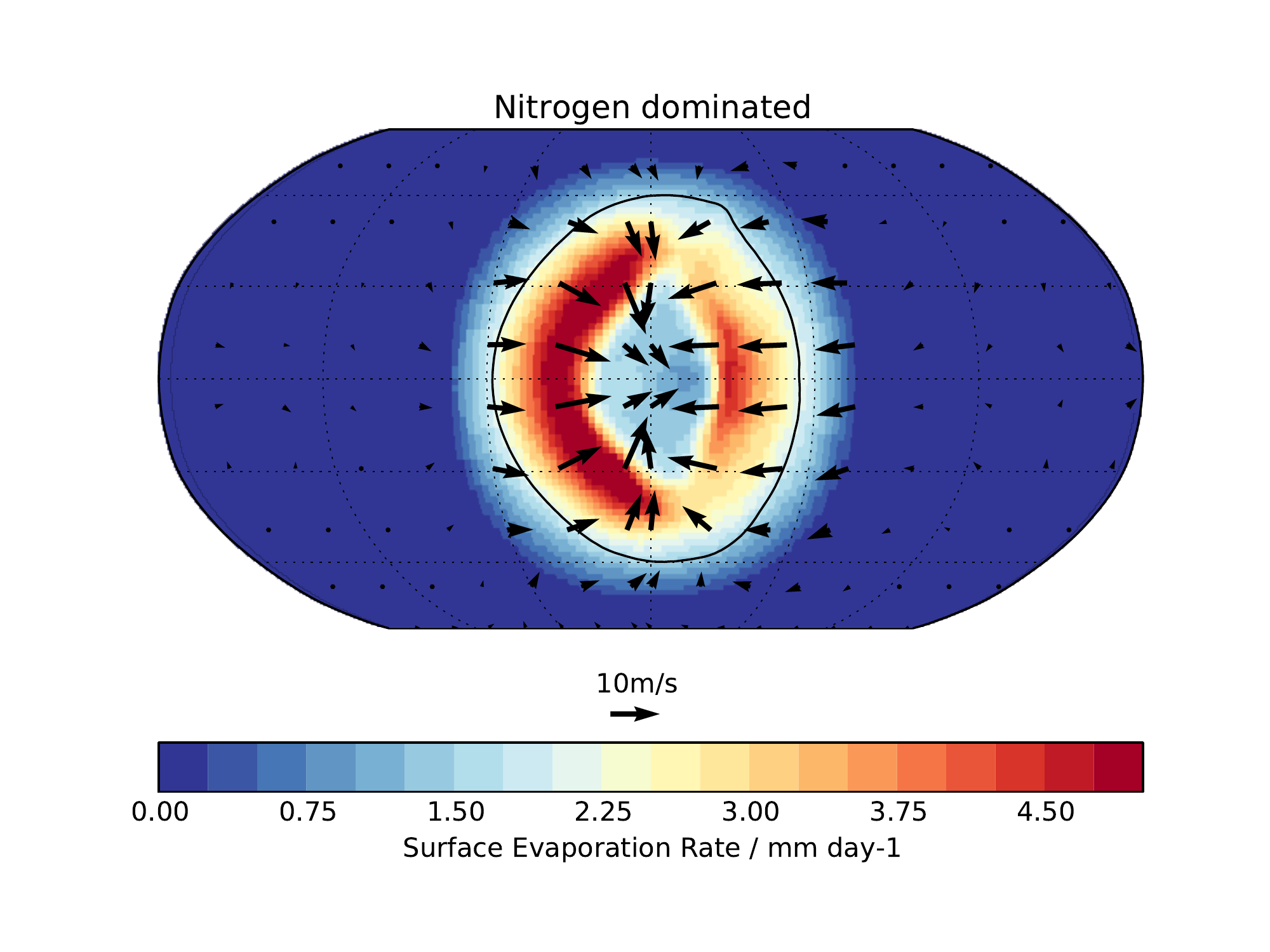}
  \caption{{ 10 orbit mean} surface precipitation and evaporation rates (coloured) with
    the mean $0\degree$C surface temperature contour and wind-vectors
    at $10$~m, for the tidally-locked nitrogen dominated atmospheric
    composition.}
  \label{fig-ppn}
\end{figure}

Figure~\ref{fig-ppn} shows the surface precipitation rate, showing
intense precipitation at, and slightly down-wind of, the sub-stellar
point on the day-side of the planet, decreasing in intensity radially
from this point. The most intense precipitation comes from deep
convection above this point, with the depth of the convection
gradually reducing with radial distance, through the congestus regime
(i.e.~convection that terminates in the mid-troposphere) and
ultimately shallow convection near the edge of the cloud layer. This
can be seen quite clearly in Figure~\ref{fig-cloud}, with cloud height
transitioning from low+medium+high, to low+medium, to low at
increasing distances from the sub-stellar point. In many ways the
transition is similar to the transition from shallow to deep
convection in the trade regions of Earth. High cloud detrained into
the anvils of convection is advected downstream by the equatorial jet,
giving rise to a distinct asymmetry in the high cloud cover. The phase
of the precipitation switches to snow in a ring around the edge of the
day-side where the temperature drops below freezing, although it is
interesting to note that the dominant phase of the precipitation is
still snow even for surface temperatures above freezing. This is due
to a combination of the time taken for the precipitation (which forms
in the ice phase) to melt at temperatures above freezing, and the fact
that near surface winds are predominantly orientated radially inwards
near the surface, which advects the snow into warmer regions.

One interesting difference from tropical circulation on Earth is that
the strong radiative heating of both the clear sky and cloud tops
effectively stabilises the upper atmosphere. This keeps the majority
of the convection quite low within the atmosphere, and only allows the
most intense events to reach the tropopause level. The surface
precipitation is therefore approximately $50$\% convective, with the
remainder being large-scale precipitation coming from the extensive
high-level cloud and driven by a large-scale ascent on the day-side of
the planet. This ascent is driven by convergence, similar to that
shown in Figure~\ref{fig-ppn}, occurring throughout the lowest few
kilometres of the atmosphere. This results in a latent heating profile
below $4$~km in Figure~\ref{fig-heat}, which is near-zero due to
averaging of intermittent convective events, which generate strong
heating, and persistent rain falling from the high-level cloud and
evaporating, cooling the air (in contrast to Earth).

Figure~\ref{fig-ppn} also shows the surface evaporation rate, and
demonstrates that the moisture source for the heaviest precipitation
is not local. The surface moisture flux is very low at the sub-stellar
point, and highest in a ring surrounding this. This inflow region to
the deep convection is where the surface winds are strongest, driving
a strong surface latent heat flux. The near surface flow moistens and
carries this water vapour into the central sub-stellar point, before
being forced upwards in the deep convection and precipitating
out. Combined with Figure~\ref{fig-ppn}, we can infer that most of the
hydrological cycle on a planet like this occurs in the region where
liquid water is present at the surface, i.e. the circulation does not
rely strongly on evaporation from regions where the surface is likely
to be frozen. Neither does the circulation transport large amounts of
water vapour into these regions, and so this configuration could be
stable for long periods if the return flow of water into the warm
region (via glaciers { or sub-surface oceans}) can match the weak
atmospheric transport out of this region.

\subsection{3:2 resonance}
\label{sub_section:resonance}

We consider now the possibility of asynchronous rotation in a 3:2
spin-orbit resonance. In this case we model an atmosphere dominated by
nitrogen, as in Section ~\ref{section:results}, and do not consider an
Earth-like composition, as the differences between the two were found
to be small for the tidally-locked case.

\begin{figure}[tbhp]
  \centering
  \includegraphics[width=\columnwidth]{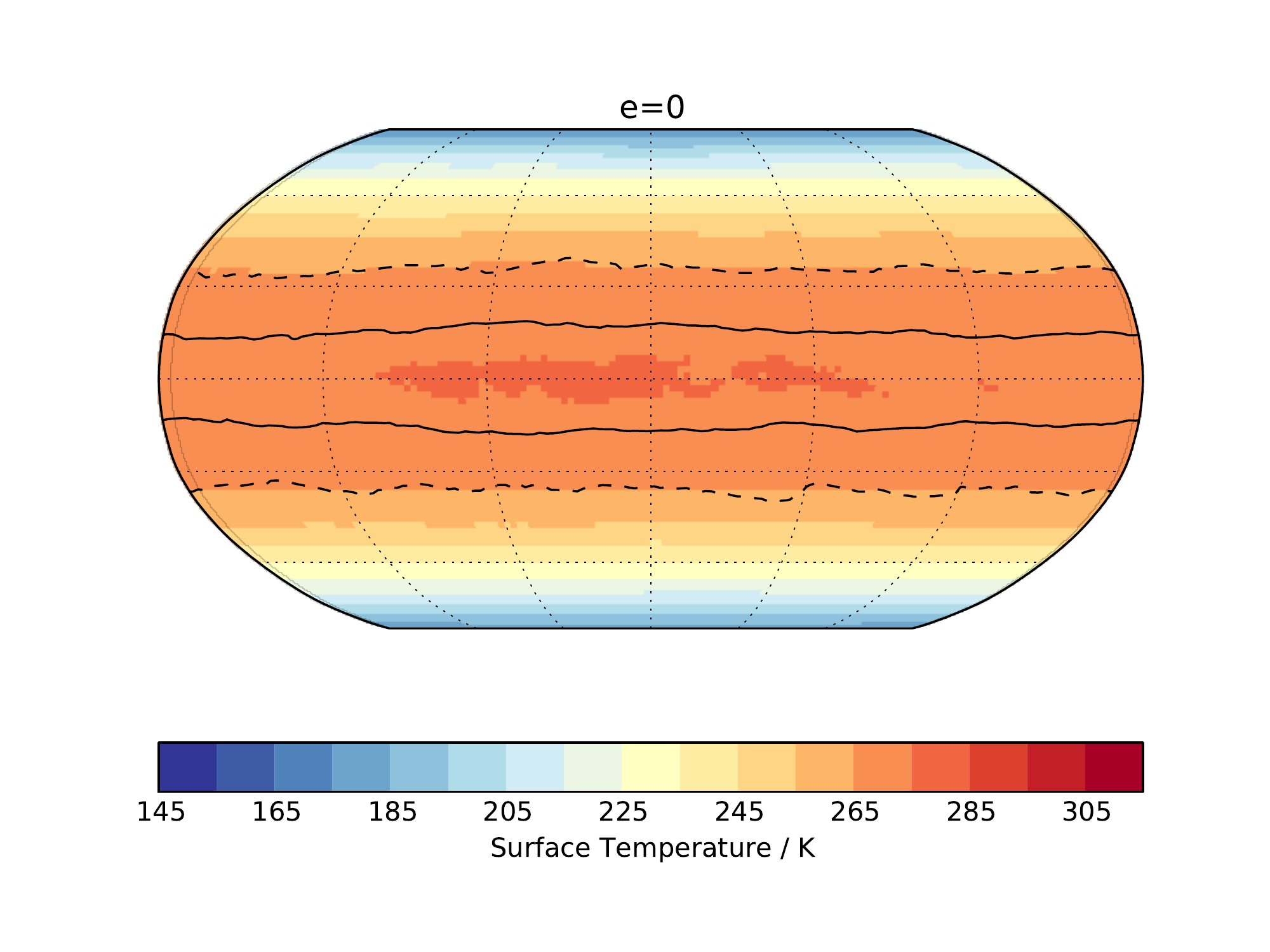}
  \includegraphics[width=\columnwidth]{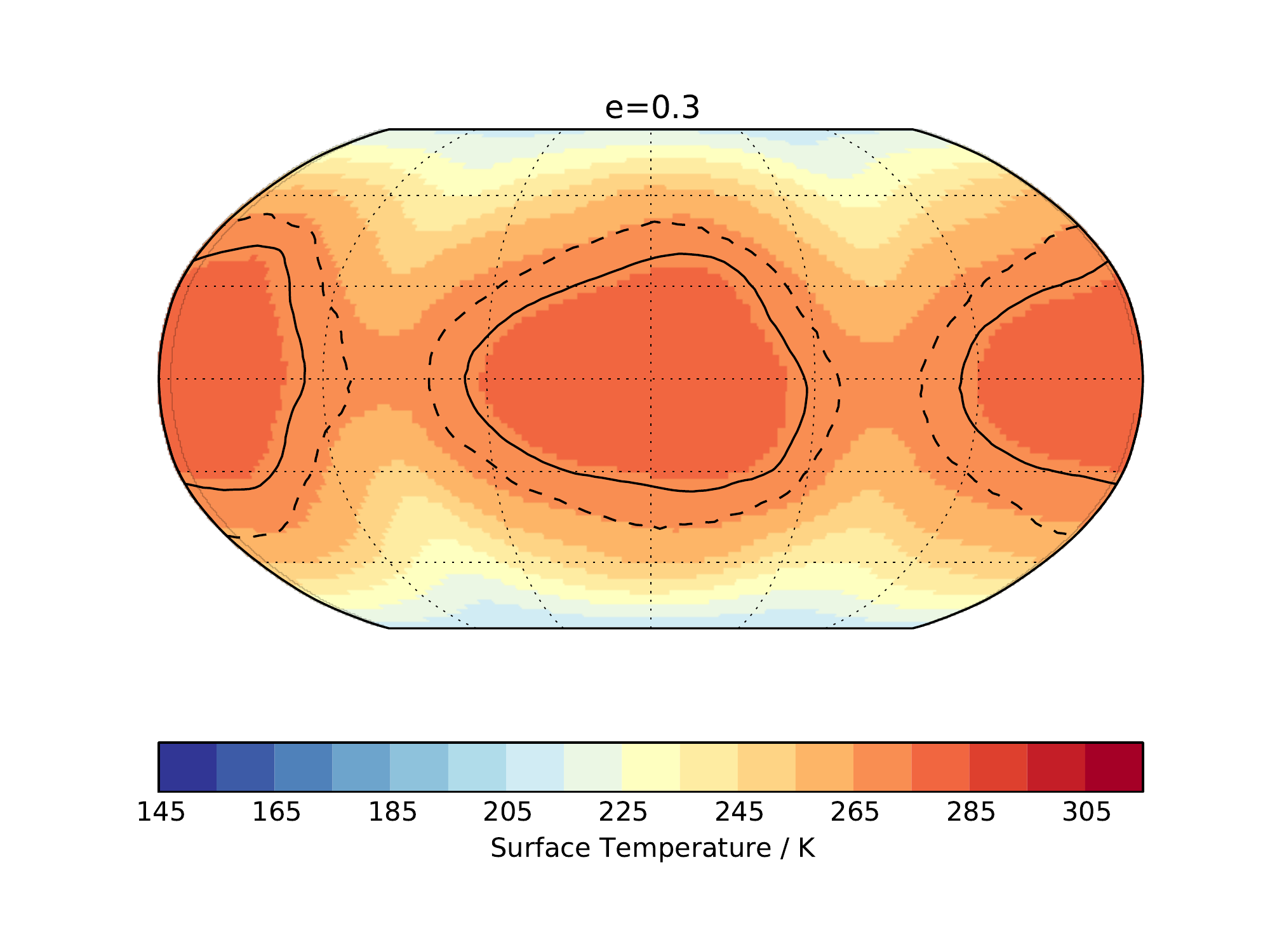}
  \caption{{ 10 orbit mean} surface temperature (coloured) with the mean (solid) and
    seasonal maximum (dashed) $0\degree$C contours, for the circular
    and eccentric orbits in a 3:2 resonance.}
  \label{fig-surft2}
\end{figure}

Figure~\ref{fig-surft2} shows the results from a circular orbit, and
{ unlike} \cite{TurLS16} { we find} that the mean surface temperature is
{ above} $0\degree$C { in a narrow equatorial band, with} seasonal maximum
temperatures above freezing extending to $35\degree$ in latitude north
and south of the equator. { There are several possible explanations for this. Firstly, the greenhouse effect may be stronger, implying that more water vapour is retained in the atmosphere in our simulations (as we know that CO$_2$ concentrations are similar). Secondly, the meridional heat transport may be weaker in our simulations, as it appears (by comparison to their figures) that our polar regions may be colder. Finally, the lack of an interactive ice-albedo at our surface may be important here. To test this, we set the surface albedo to $0.27$ everywhere, to be representative of an ice covered surface, based on the mean spectral albedo of the surface ice/snow cover calculated by \cite{TurLS16}. We find that in this state, the mean surface temperature does fall below $0\degree$C everywhere (not shown), although seasonal maximums above freezing are still retained. Therefore, although the mean temperature of the planet is higher in our simulations, it is still likely that this configuration would fall into a snowball state.}

The chance of a planet existing in a
resonant orbit with zero eccentricity is small \cite[]{GolP66} yet if
ProC B is the only planet in the system, the eccentricity excited by
$\alpha$~Centauri alone is likely to be $\approx 0.1$
\cite[]{RibBS16}. Current observations can not exclude a further
planet(s) orbiting exterior to ProC B, and are consistent with an
eccentricity as large as $0.35$ \citep{AngABB16}, { with the most likely
estimate $0.25$ \cite[]{Bro17}}. Therefore we have
run a range of simulations assuming a 3:2 resonant orbit but with
eccentricities varying from zero to $0.3$, and focus discussion on the
most eccentric case. { With increasing eccentricity, the region where the mean surface temperature is above freezing becomes concentrated in two increasingly large patches,}
corresponding to the side of the planet
which is facing the star at periastron on each orbit. Therefore,
permanent liquid water could exist at the planet surface, and the
potential for the planet to fall into a snowball state is greatly
reduced.

\begin{figure}[tbhp]
  \centering
  \includegraphics[width=\columnwidth]{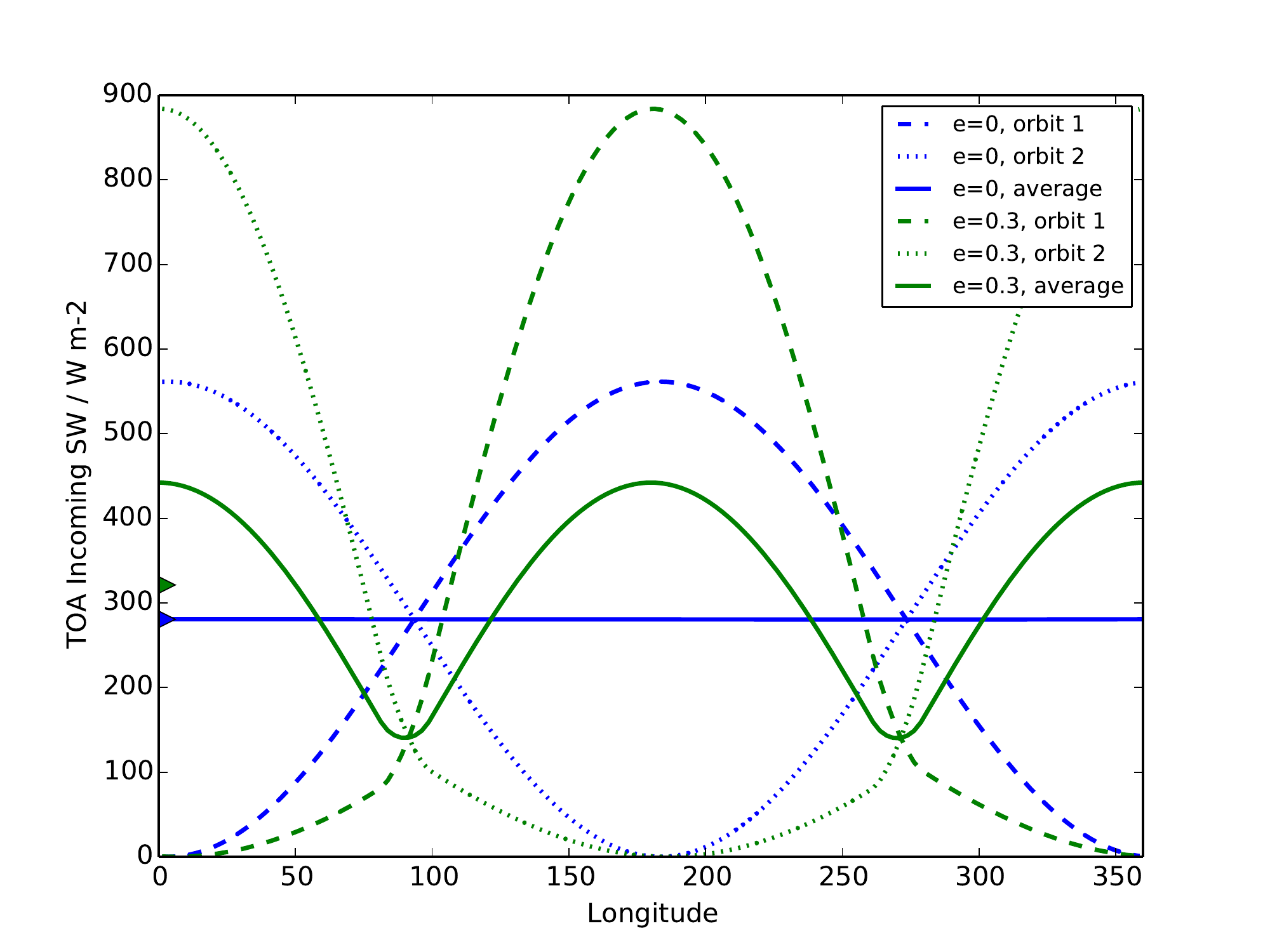}
  \caption{Equatorial cross-section of the top-of-atmosphere incoming
    shortwave radiation, meaned over a single orbit, showing two
    consecutive orbits for the circular and eccentric configurations
    (dashed and dotted) in a 3:2 resonance. Also shown is the mean of
    these two consecutive orbits (solid), and the zonal mean of this
    (filled triangles).}
  \label{fig-swin}
\end{figure}

In an eccentric orbit the stellar heating is concentrated in two
hot-spots on opposite sides of the planet { \cite[]{Dob15}}, leading to large regions of
the surface which are warmer than their
surroundings. Figure~\ref{fig-swin} shows how the incoming
top-of-atmosphere shortwave radiation varies between orbits for the
circular and eccentric configurations. The increase in radiation as
the eccentric orbit approaches periastron is much greater than the
decrease in radiation as it approaches { apoastron}, resulting in a
significant increase in the mean stellar flux over large regions of
the planet. This, combined with the fact that the total equatorial
radiation is increased by $40$~W~m$^{-2}$ (the global mean is closer
to $30$~W~m$^{-2}$), keeps the hot-spots well above freezing, with
mean temperatures { above} $280$~K and seasonal maximums of $295$~K. The
global mean temperature however only rises by $4$~K, which for an
effective increase in stellar flux of $120$~W~m$^{-2}$ ($30$
multiplied by the surface-area to disc ratio of $4$) implies an even
lower sensitivity of this orbital state to changes in the stellar flux
than in the tidally-locked case. We find that the sensitivity to the
stellar flux changes for a 3:2 resonance in a circular orbit is
approximately equal to the tidally-locked case ($5$~K of warming for
$70$~W~m$^{-2}$ additional flux), implying that the lower sensitivity
is due to the eccentricity of the orbit -- the high cloud cover formed
over the hot-spot regions (Fig.~\ref{fig-cloud2}) increases the
reflected shortwave radiation, increasing the planetary albedo. { This shows, similar to \cite{BolLL16}, that the mean flux approximation is poor for this eccentric orbit.}

\begin{figure}[tbhp]
  \centering
  \includegraphics[width=\columnwidth]{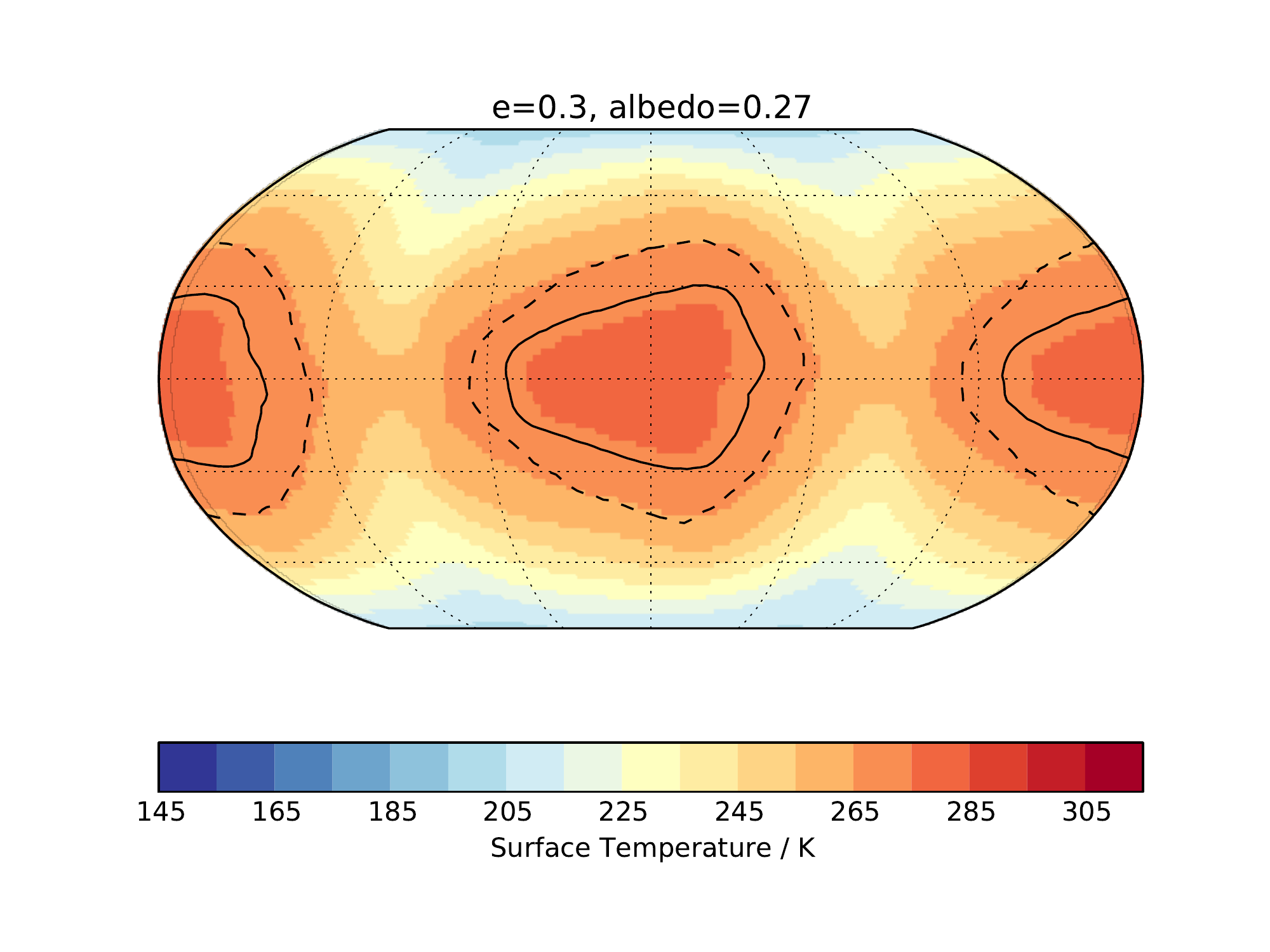}
  \caption{{ 10 orbit mean} surface temperature (coloured) with the mean (solid) and
    seasonal maximum (dashed) $0\degree$C contours, for the eccentric
    orbit in a 3:2 resonance with surface albedo representative of an
    ice/snow covered surface.}
  \label{fig-surft3}
\end{figure}

\begin{figure}[tbhp]
  \centering
  \includegraphics[width=\columnwidth]{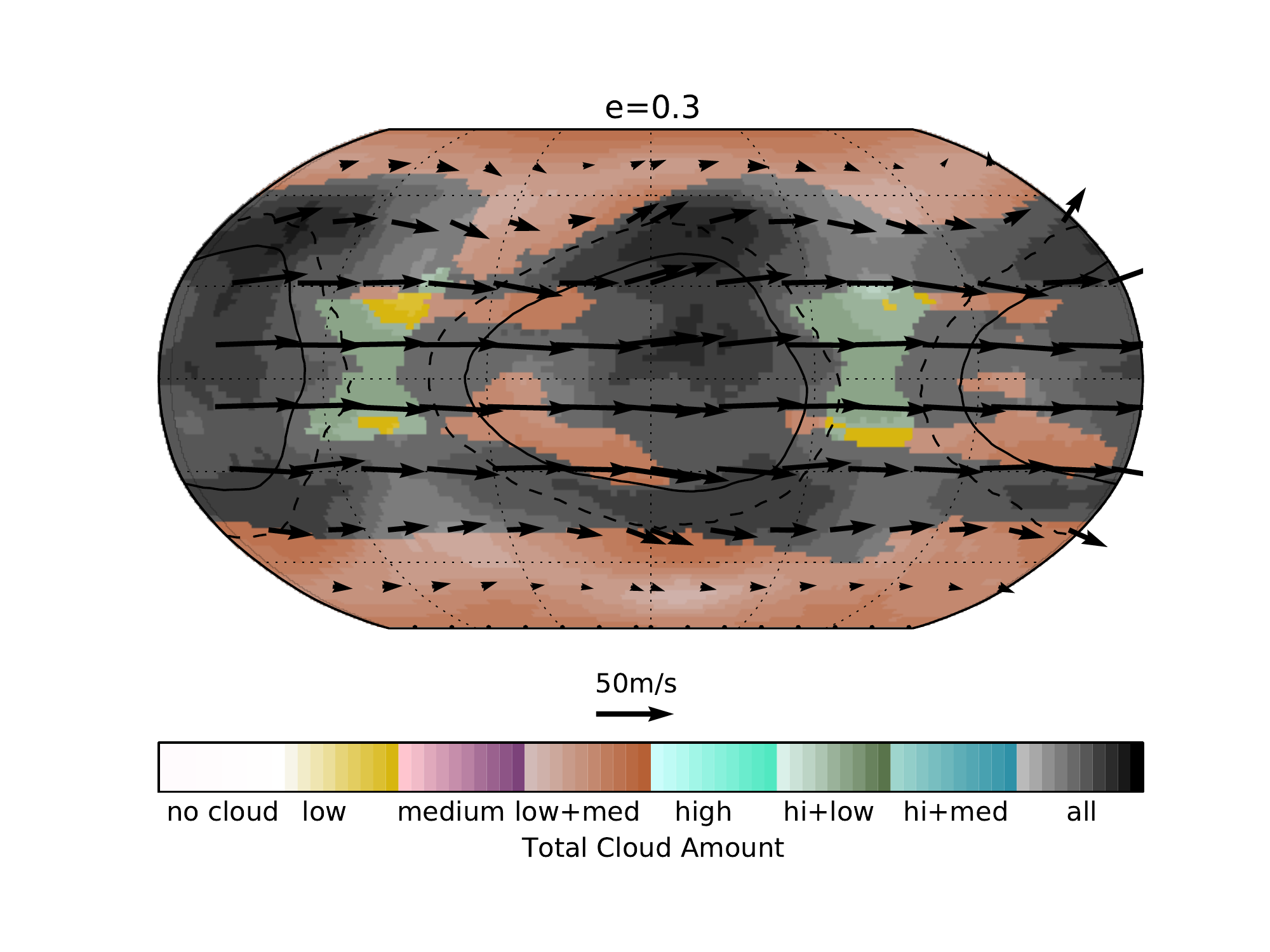}
  \caption{{ 10 orbit mean} cloud cover (coloured) with the mean (solid) and seasonal
    maximum (dashed) $0\degree$C surface temperature contours and
    wind-vectors at $8.5$~km, for the eccentric orbit in a 3:2
    resonance. Low cloud is below $2$~km and high cloud is above
    $5.5$~km, { see Fig.~\ref{fig-cloud} for more details}.}
  \label{fig-cloud2}
\end{figure}

To test the possibility of the planet falling into a snowball state in
this orbital configuration, we { again} set the surface albedo to be $0.27$
everywhere, { to represent a snow/ice covered surface.}
This represents the most extreme scenario possible, as
it would imply that any liquid water at the surface has managed to
freeze during the night, which only lasts $12$ Earth
days. Figure~\ref{fig-surft3} shows that even in this case, the mean
surface temperature remains above zero { (and in fact the minimum only just reaches freezing)}, implying that the chance of
persistent ice formation in these regions is small and the planet is
unlikely to snowball. Additional tests with intermediate values of
orbital eccentricity allow us to estimate that an eccentricity of
$\approx 0.1$ would be required to maintain liquid water at the surface
and prevent this configuration falling into a snowball state. The
intermediate eccentricity simulations display many features of the
most eccentric orbit presented here, i.e.~the formation of hot-spot
regions, but their strength obviously increases with increasing
eccentricity.

GCM studies of resonant orbits with eccentricity appear to be rare,
and therefore we document here what this climate might look like in
the most eccentric case. In many respects, it appears similar to the
tidally-locked case presented in Section~\ref{sec-fixed}, except now
with two hot-spots on opposite sides of the planet and a much reduced
planetary area in which water would be frozen. There are no
significant cold-traps, with the polar regions being the coldest area
with surface temperatures { just above} $200$~K, not too dissimilar from Earth
(surface temperatures in Antarctica are typically $210$~K). As this
simulation contains no ocean circulation, we speculate that if
this were included, it could transport more heat away from the
hot-spots and further warm the cold regions.

\begin{figure}[tbhp]
  \centering
  \includegraphics[width=\columnwidth]{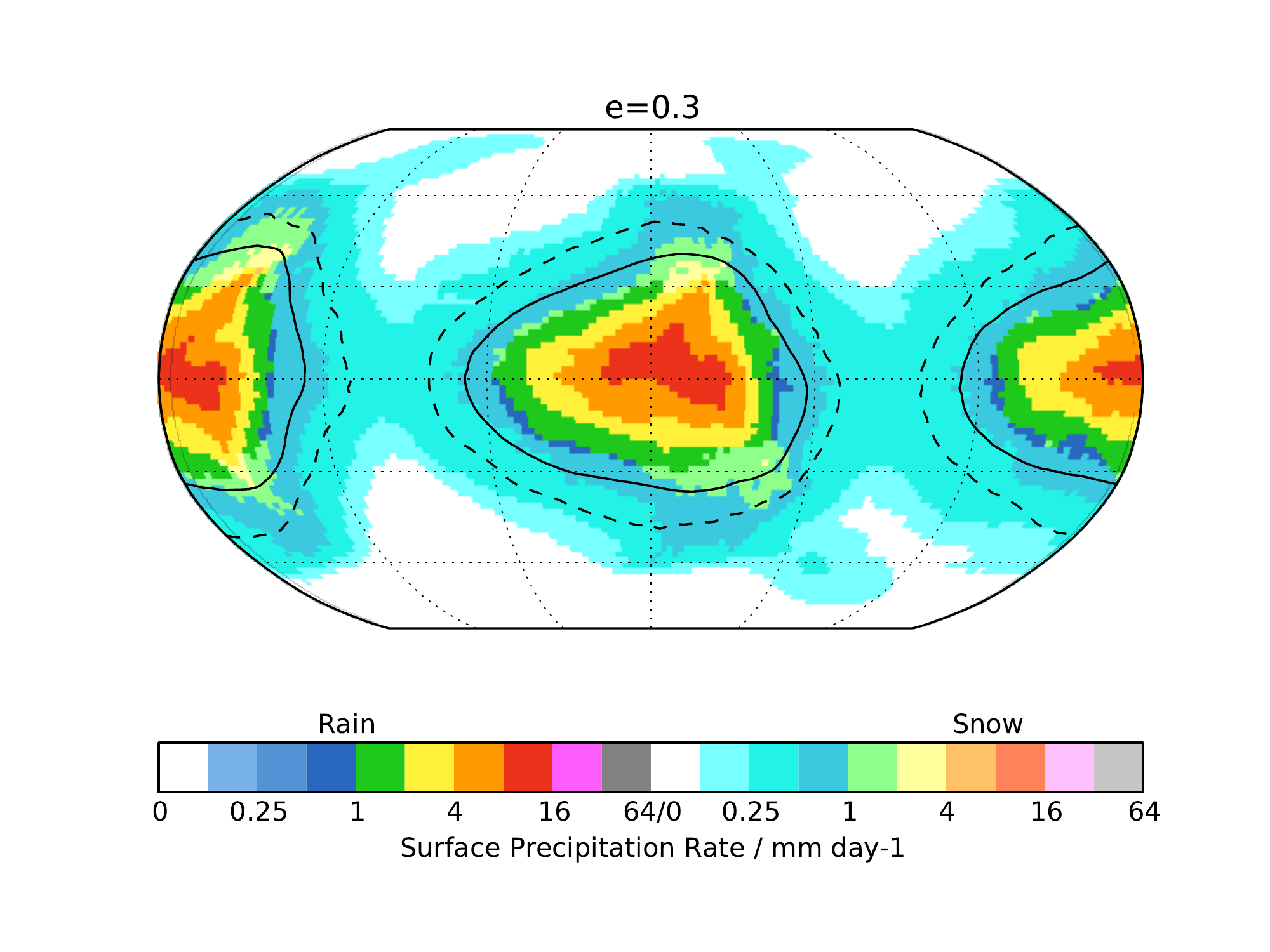}
  \includegraphics[width=\columnwidth]{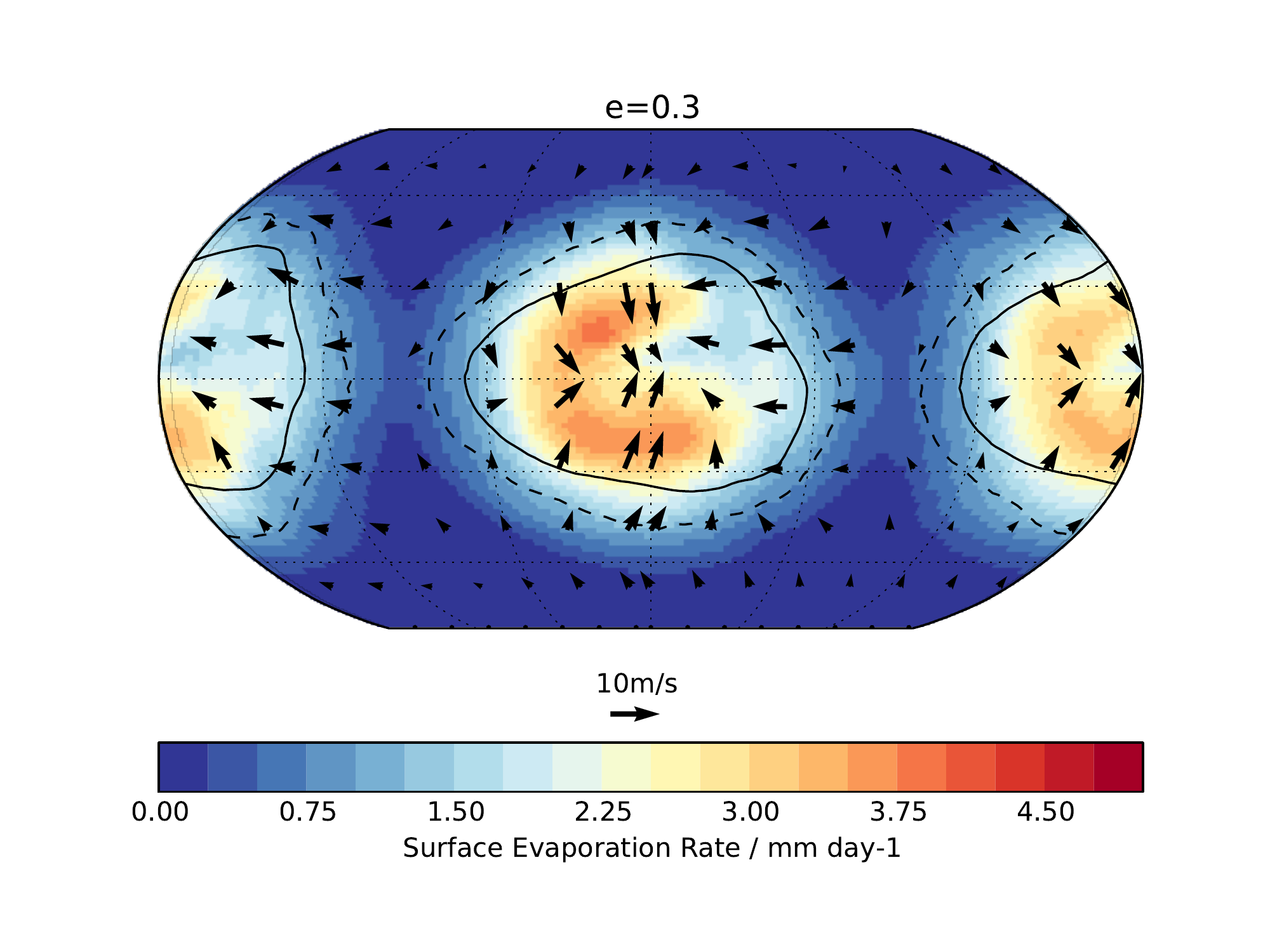}
  \caption{{ 10 orbit mean} surface precipitation and evaporation rates (coloured) with
    the mean (solid) and seasonal maximum (dashed) $0\degree$C surface
    temperature contours and wind-vectors at $10$~m, for the eccentric
    orbit in a 3:2 resonance.}
  \label{fig-ppn2}
\end{figure}

Figures~\ref{fig-cloud2} and \ref{fig-ppn2} show that the hot-spots
are dominated by deep convective cloud with heavy precipitation. The
upper-level circulation appears to be dominated by a zonal jet
covering most of the planet, and similar to the tidally-locked case,
this acts to advect the convective anvils downstream into the cold
regions of the planet, { almost completely encircling the planet}, and maintain a horizontally uniform temperature
in the free troposphere of the planet (not shown). Most of the planet
is covered in low and mid-level cloud, apart from sub-tropical regions
downstream of each hot-spot where there is only low-level cloud,
similar to persistent stratocumulus decks on Earth. These cloud decks
form in the regions of large-scale subsidence, compensating for the
large-scale ascent which occurs in the hot-spot regions.

The cloud is thick enough to precipitate around the entire equatorial
belt, and reaches $60\degree$ north/south at the central longitude of
the hot-spots. Similar to the tidally-locked case, the heaviest
precipitation is downstream of the peak stellar irradiation, due to
the strong zonal winds. The winds also shift polewards at this
location, which creates two limbs of enhanced precipitation stretching
polewards downstream of the hot-spot. The low level flow is generally
equatorwards (Fig.~\ref{fig-ppn2}), and this is strongest in the warm
regions, leaving weak winds in the colder subtropical regions, ideal
for the formation of non-precipitating low cloud.

Figure~\ref{fig-ppn2} also shows the surface evaporation rate, which
unlike the tidally-locked case is not strongly confined to a region
surrounding the deepest convection, but instead appears quite local to
the precipitation. It is predominantly to the upstream side of the
heaviest precipitation, and this is due to the lower cloud cover in
this region allowing more stellar radiation to reach the surface. The
hydrological cycle of each hot-spot appears reasonably self contained,
with the possibility that there will only be limited exchanges of
water between the opposing sides of the planet. Similar to the
tidally-locked case, the hydrological cycle is also largely confined
within regions where surface liquid water is present, suggesting that
this configuration could be stable for long periods.

\begin{figure}[tbhp]
  \centering
  \includegraphics[width=\columnwidth]{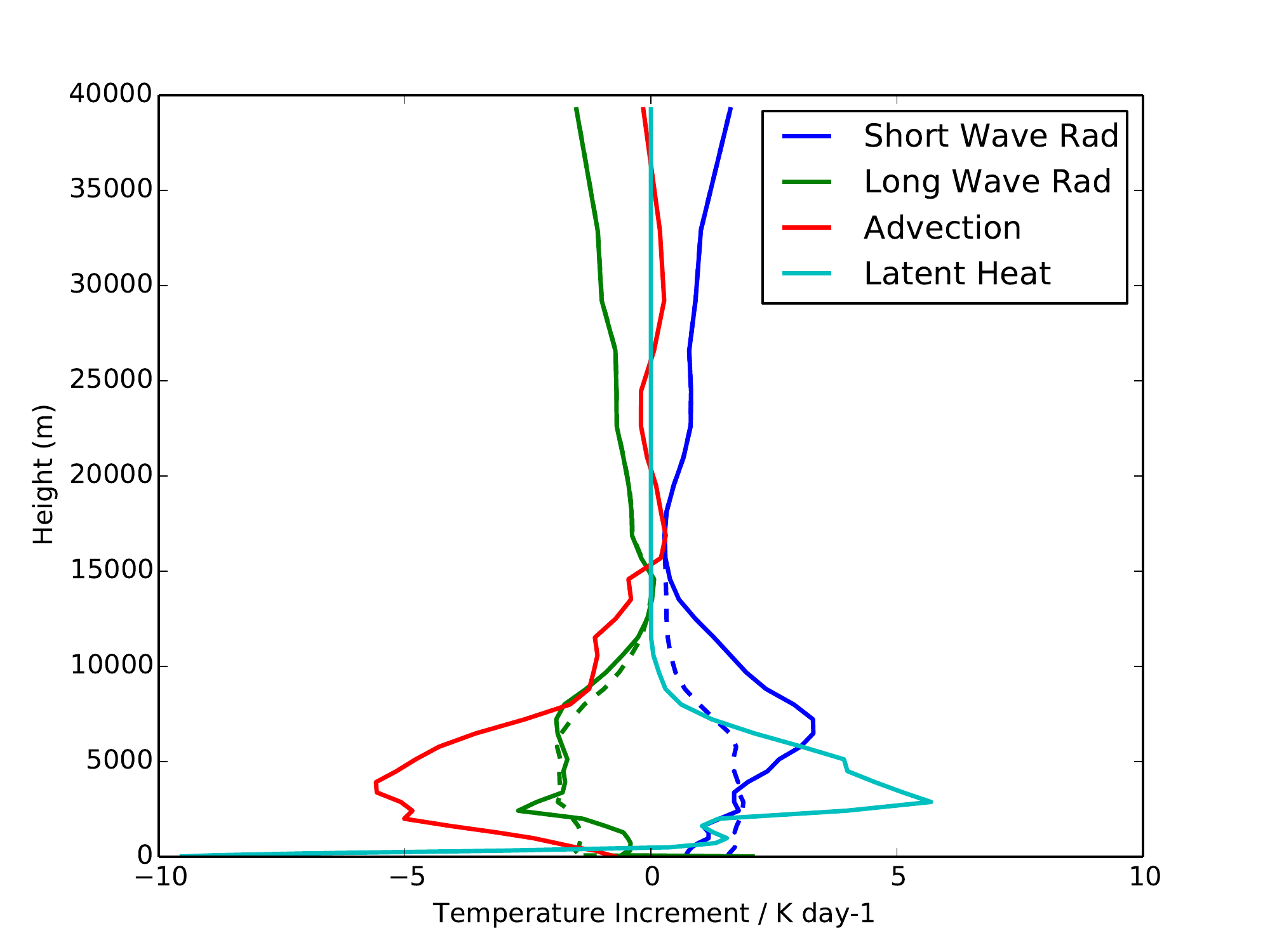}
  \caption{{ 10 orbit mean} heating profiles from the main physical processes, at the
    centre of one of the hot-spot regions of the planet in an
    eccentric orbit with 3:2 resonance. The dashed lines show the
    clear-sky heating from the radiation components.}
  \label{fig-heat2}
\end{figure}

Whilst there are many similarities in the resultant climate between
the tidally-locked case and the hot-spots on the 3:2 eccentric case,
the mechanisms which create them do show some differences. The 3:2
case is more similar to Earth in its heating profiles
(Fig.~\ref{fig-heat2}), with latent heating due to convection
dominating over shortwave heating throughout the lower-to-mid troposphere. The magnitude of the shortwave heating is much lower than in
the tidally-locked case, although still significantly higher than on
Earth \cite[]{McFMA07}, and this does not act to stabilise the upper
troposphere and suppress the convection. The resulting climate
therefore has deep convection mixing throughout the troposphere during
the day, and the majority of the precipitation is convective in this
simulation. It therefore presents a somewhat intermediate solution
between the tidally-locked case and Earth. The fact that the whole
planet is irradiated at some point during the orbit means there is no
stratospheric transport of heat from day to night, with the
stratospheric heating and cooling being entirely radiative in nature.

\section{Spectra and Phase Curves}
\label{section:spectra}

We have produced individual, high-resolution reflection (shortwave)
and emission (longwave) spectra, as well as reflection/emission as a
function of { orbital phase angle} within wavelength
`bins'. \cite{TurLS16} include a full discussion on the possibility of
detecting this planet with current and upcoming instrumentation, so we
do not repeat this here. Instead, we present our simulated emissions
and discuss their key features, and differences with \cite{TurLS16}.

We obtain the top-of-atmosphere flux directly from our GCM simulation,
as done by \citet{TurLS16}, where the wavelength resolution is defined
by the radiative transfer calculation in the GCM. The radiative
transfer calculation used in our GCM, and that of \cite{TurLS16},
adopts a correlated-$k$ approach, effectively dividing the radiation
calculation into bands. The bands themselves, and details of the
correlated-$k$ calculation are essentially set to optimise the speed
of the calculation while preserving an accurate heating rate \cite[see
discussion in][]{AmuBT14,AmuMB16}. Therefore, spectral resolution is
sacrificed.

To mitigate this effect, we run the model for { a single orbit} with a
much greater number of bands (260 shortwave and 300 longwave), which
greatly increases the spectral resolution. { The top-of-atmosphere flux is output every two hours, or approximately 144 times per orbit for ProC B.} From these simulations we
compute the emission and reflection spectra, as well as the
reflection/emission as a function of { time and orbital phase. The outgoing flux at the top of the model atmosphere is translated to the flux seen at a distant observer by taking the component of the radiance (assumed to be isotropic) in the direction of the observer and then summing over the solid angle subtended by each grid point over the planetary disc}.

\begin{figure*}[tbhp]
  \centering
  \includegraphics[width=\columnwidth]{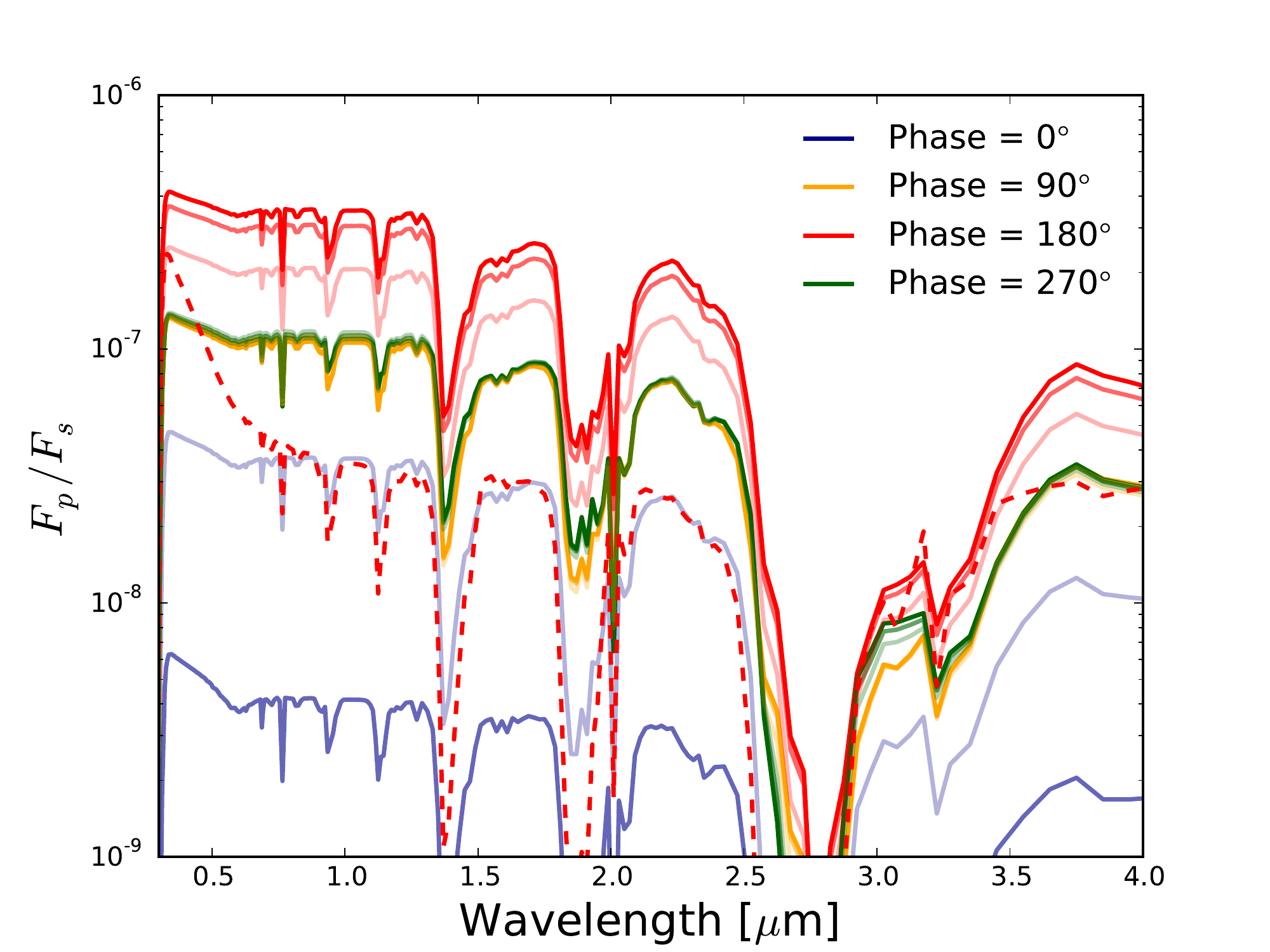}
  \includegraphics[width=\columnwidth]{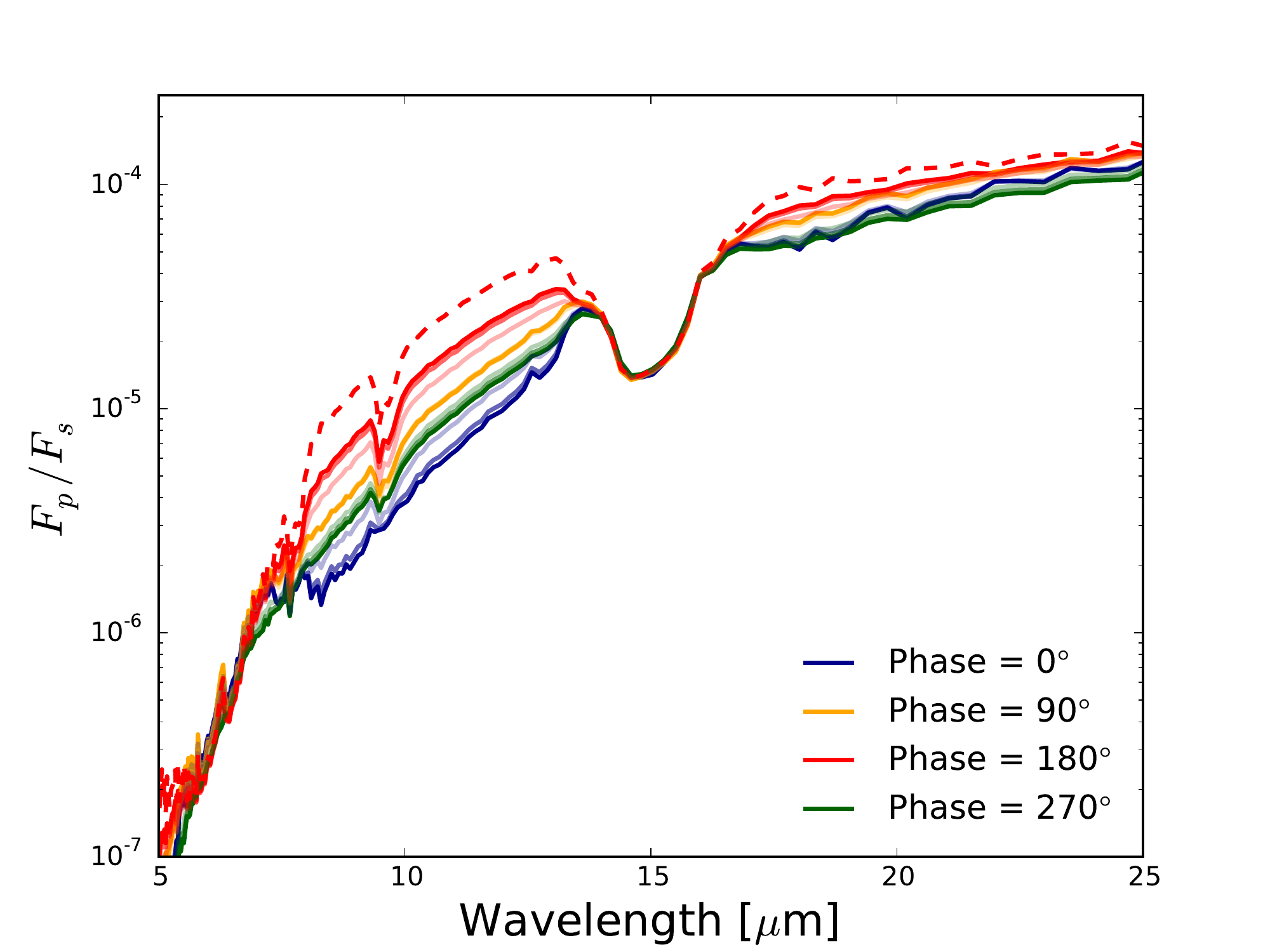}
  \caption{Reflectance ({\it left}) and emission ({\it
      right}) spectra for the tidally-locked Earth-like case at four
    points in the planets orbit. Spectra for observer inclinations of
    90$\degree$, 60$\degree$ and 30$\degree$ are shown in
    progressively fainter lines. For comparison, we include the
    clear-sky flux for the phase = 180$\degree$ spectrum with an
    observer inclination of 90$\degree$ (dashed), which highlights the
    importance of the role of clouds in the shortwave spectrum. The
    sharp drop-off in the flux ratio at short wavelengths is due to
    ozone absorption}
  \label{el_tl_emission}
\end{figure*}

\begin{figure*}[tbhp]
  \centering
  \includegraphics[width=\columnwidth]{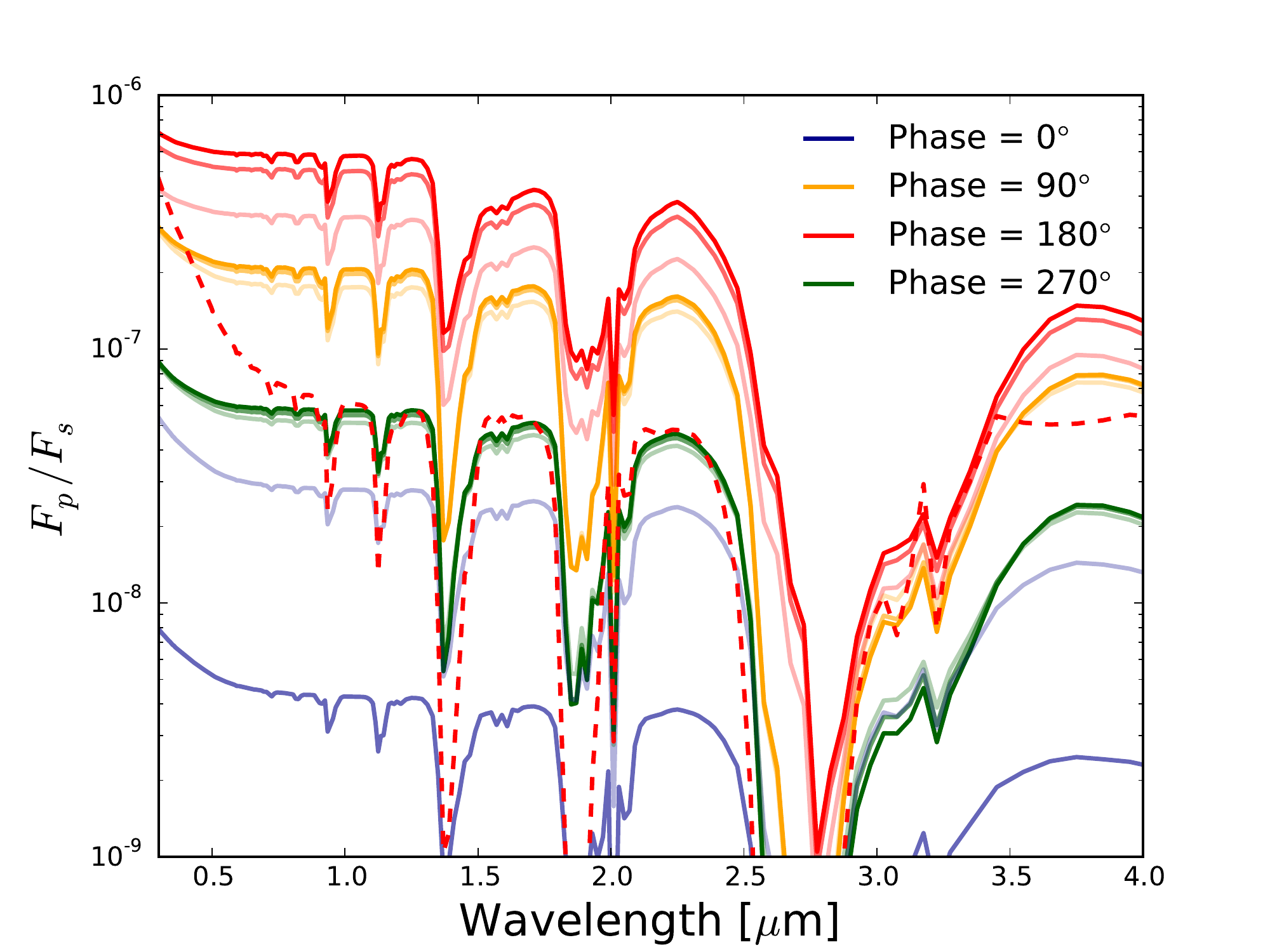}
  \includegraphics[width=\columnwidth]{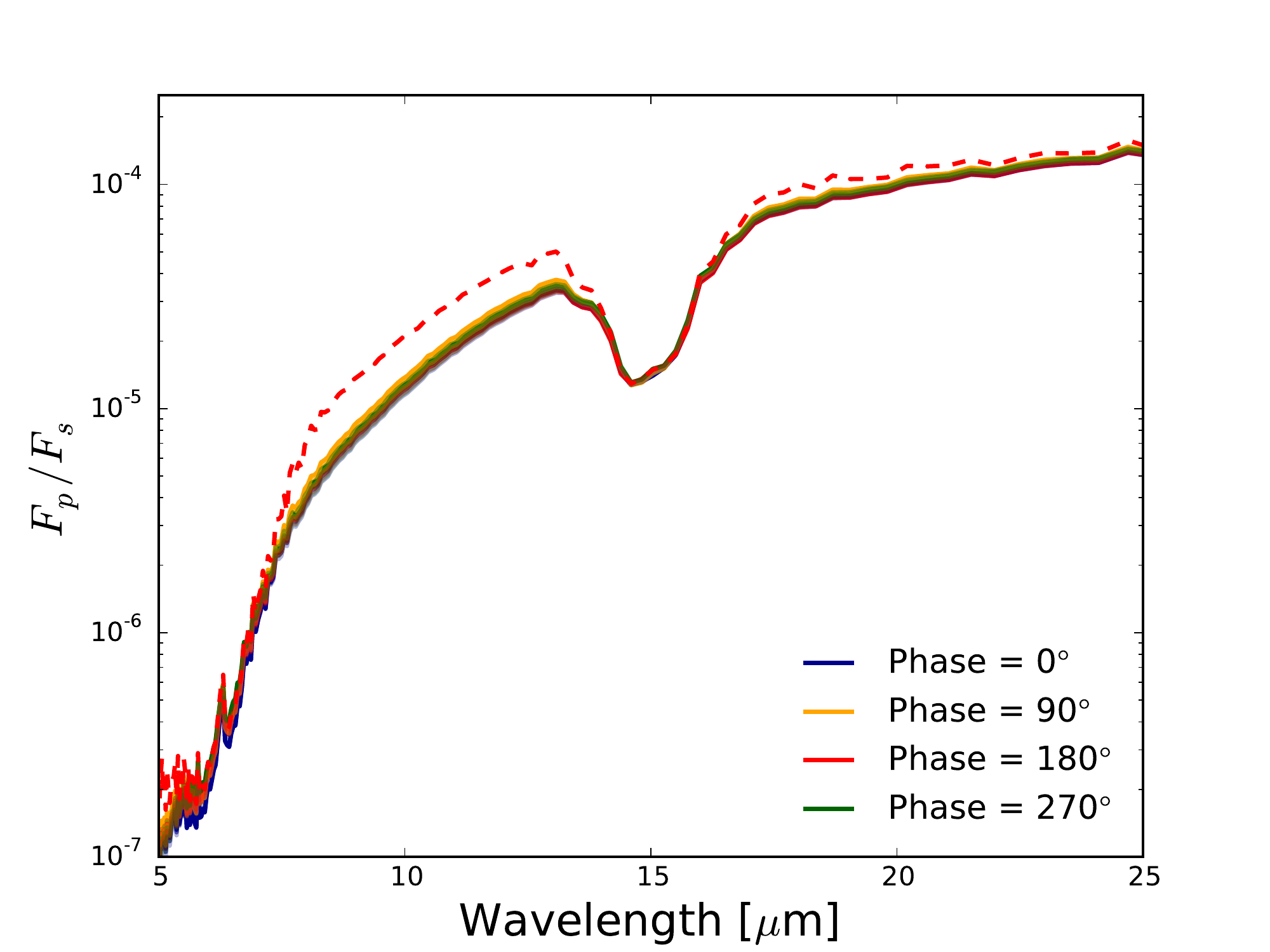}
  \caption{As Figure~\ref{el_tl_emission}, but for the 3:2 resonance
    nitrogen-dominated case, and we show the clear-sky flux for phase
    = 180$\degree$.}
  \label{n2_3_2_emission}
\end{figure*}

Figure \ref{el_tl_emission} shows the reflection (shortwave) and
emission (longwave) planet-star flux ratio for the tidally-locked case
with an Earth-like atmosphere. The contrast between planet and star
($F_p/F_s$) is shown as a function of wavelength (in $\mu$m) for { a
range of orbital phases and inclinations ($i$)}. We follow the approach of
\citet{TurLS16}, where an inclination $i$ = 90$\degree$ represents the
case where the observer is oriented perpendicular to the orbital
axis. Additionally, an example for the ``clear-sky'' emission is shown
(dashed line), ignoring the radiative effect of the cloud in the
simulated observable. These figures are comparable to their
counterparts in \cite{TurLS16} (Figures 8 and 12). We have separated
the long and shortwave flux \citep[meaning our contrast is
underestimated for the shortwave at wavelengths longer than 3.5$\mu$m,
as is the case in][]{TurLS16}, and adopted a radius of
1.1~R$_\oplus$. The differences will then be caused by the direct
differences in the top of atmosphere fluxes obtained in our
simulations, and the resolution of our emission calculation.

In the shortwave case, our spectrum generally compares well with that
of \citet{TurLS16}, showing similar trends and features, particularly
the absorption features from water and CO$_2$. However, we find an
overall larger $F_p/F_s$ ratio (e.g. by a factor of~2 between 2.0 and
2.5 $\mu$m), which is likely to be the result of subtle differences in
the quantity and distribution of clouds, which have a significant
influence on the shortwave reflection, as shown by the clear-sky
spectrum, also shown in Figure~\ref{el_tl_emission}. Our inclusion of
the full complement of Earth's trace gases along with increased
resolution reveals more spectral features, especially at short
wavelengths, but the overall shape and magnitude of the contrasts
compare well to \cite{TurLS16}. Our inclusion of oxygen leads to the absorption feature
at 0.76 $\mu$m, and an ozone layer to the absorption at ultra-violet
wavelengths. As discussed previously, the presence of an ozone layer
in the atmosphere of ProC B is very uncertain. However, our aim here
is to explore how a truly Earth-like atmosphere would respond to the
irradiation received by ProC B. Comparing the shortwave contrast from our outputs at
$\phi$=180$\degree$ with (solid line) and without (dashed line) clouds
we can see that the cloud acts to increase the shortwave contrast due
to scattering and slightly `mute' the absorption features.

Figure~\ref{el_tl_emission} also shows a reasonable agreement of the
overall shape and magnitude of our longwave flux ratio with that of
\citet{TurLS16}. As for the shortwave case our increased resolution
reveals additional features in the spectrum. { The main difference here is the absorption feature at $9.6\mu$m due to the presence of ozone, which is not included in the model of \citet{TurLS16}.}

\begin{figure}[tbhp]
  \centering
  \includegraphics[width=0.8\columnwidth]{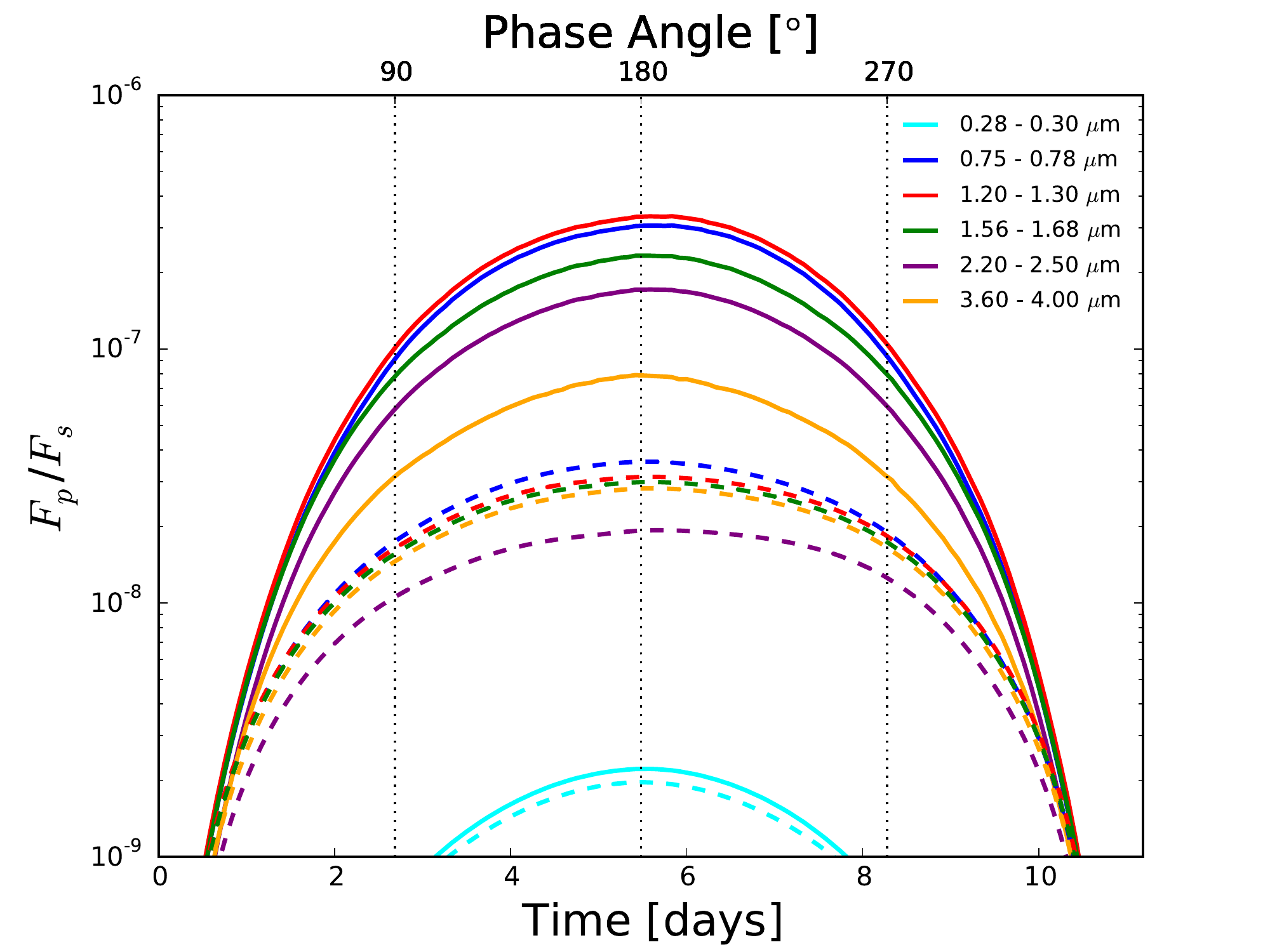}
  \includegraphics[width=0.8\columnwidth]{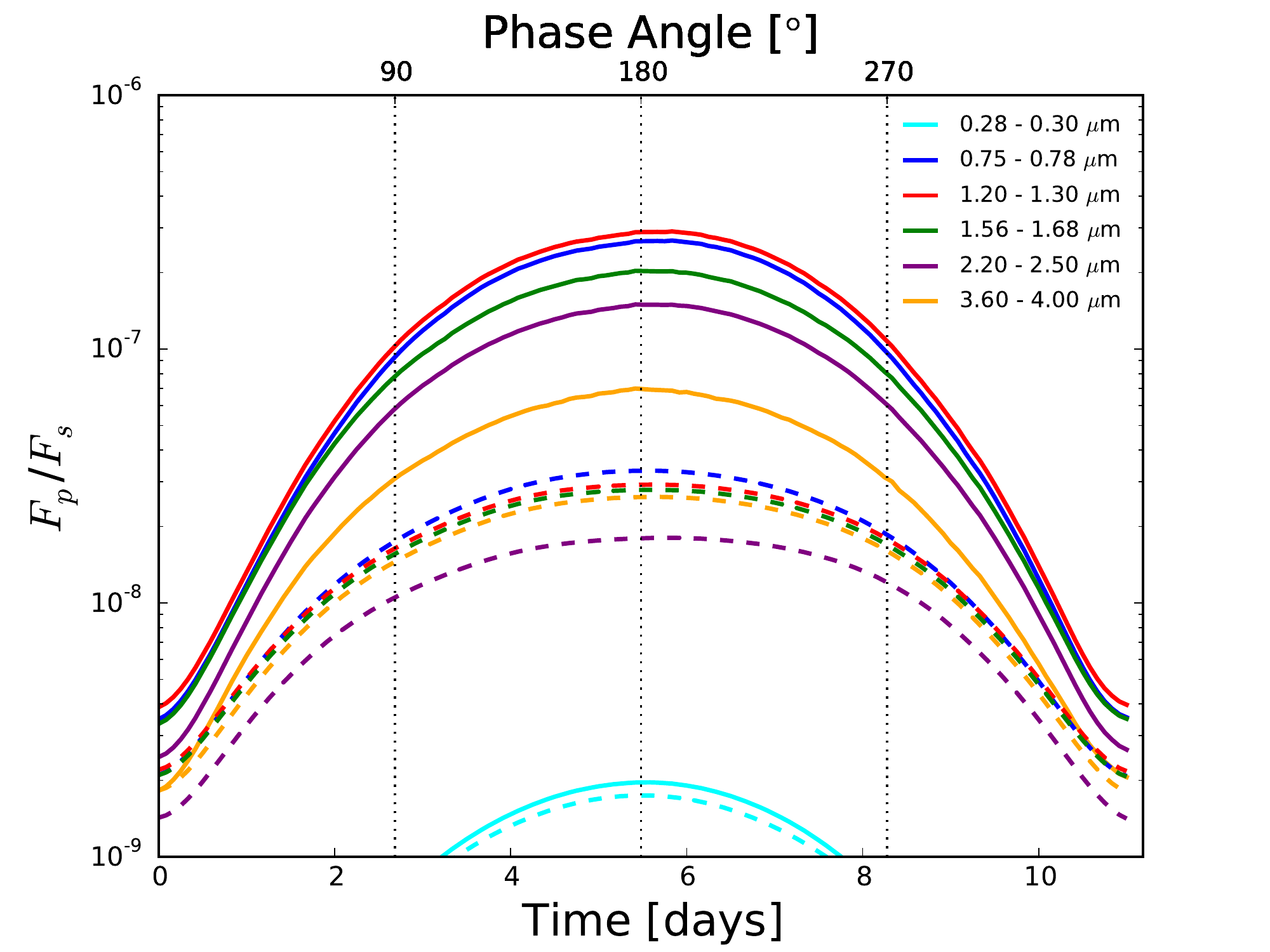}
  \includegraphics[width=0.8\columnwidth]{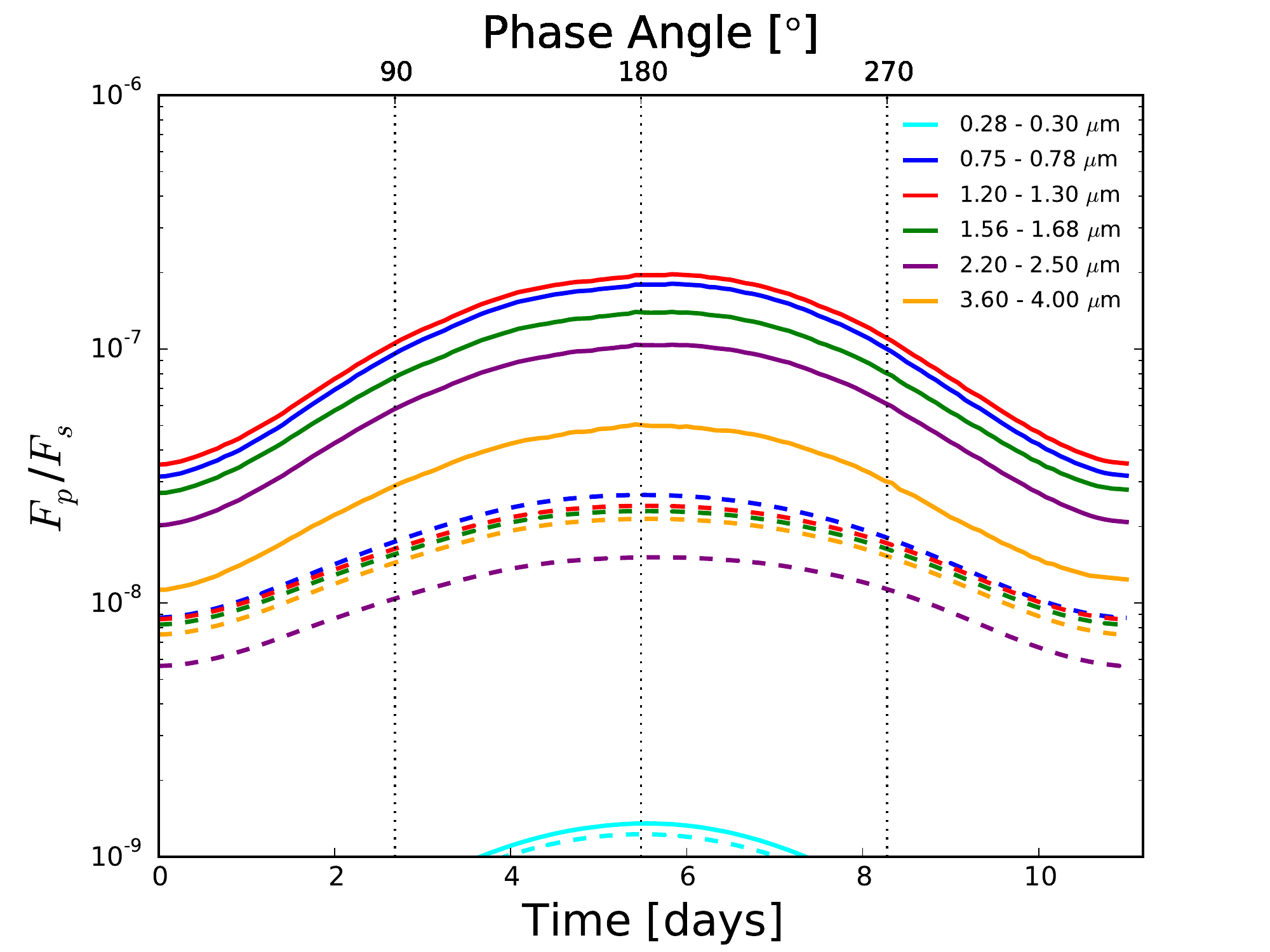}
  \includegraphics[width=0.8\columnwidth]{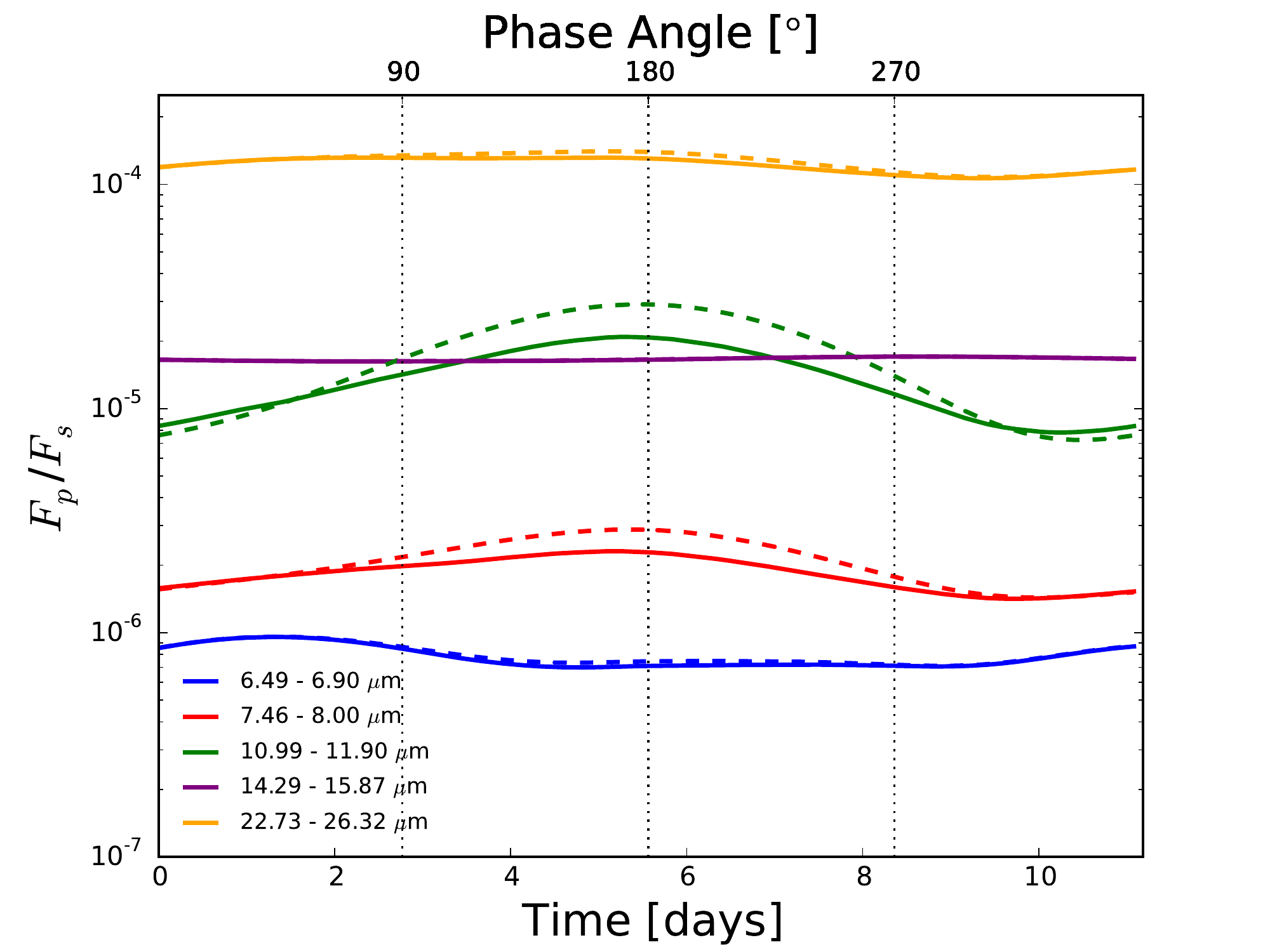}
  \caption{Reflectance phase curves for observer inclinations of ({\it
      from top to second from bottom}) 90$\degree$, 60$\degree$ and
    30$\degree$ and the emission phase curve for an observer
    inclination of 60$\degree$ ({\it bottom}) for the tidally-locked
    Earth-like case. Phase curves are shown for wavelength ranges
    closely matching those of \citet{TurLS16} for
    comparison. Clear-sky fluxes are represented in dashed lines to
    highlight the role of clouds, which generally increase the
    shortwave planetary flux but have much more subtle effects for the
    longwave flux.  The emission phase curves show very little
    variation with observer inclination, hence we only show the $i$ =
    60$\degree$ case for brevity. Note that we find a very small flux
    ratio in the 0.28-0.30$\mu$m region,
    due to ozone absorption.}
  \label{el_tl_phase}
\end{figure}

\begin{figure}[tbhp]
  \centering
  \includegraphics[width=0.8\columnwidth]{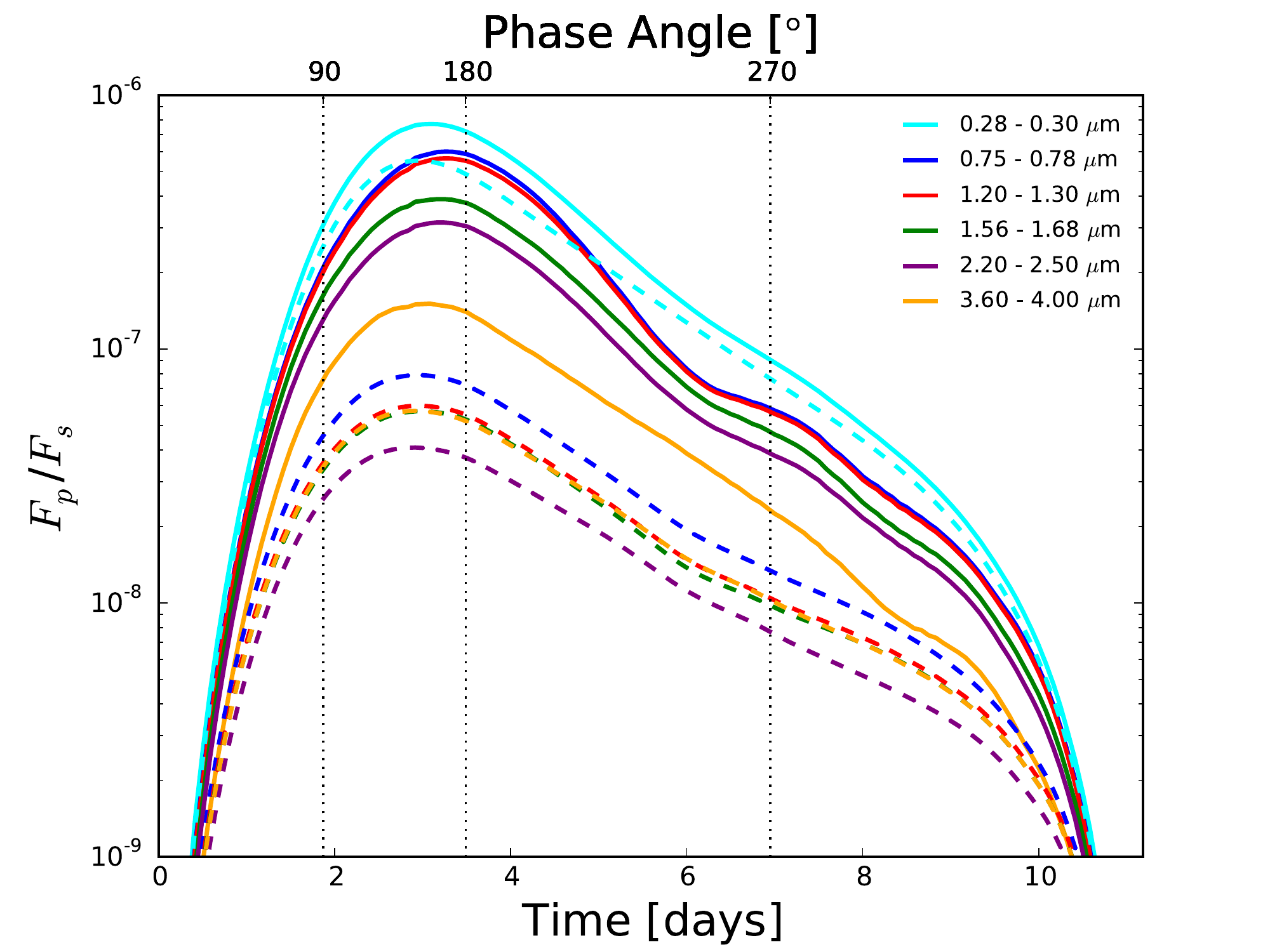}
  \includegraphics[width=0.8\columnwidth]{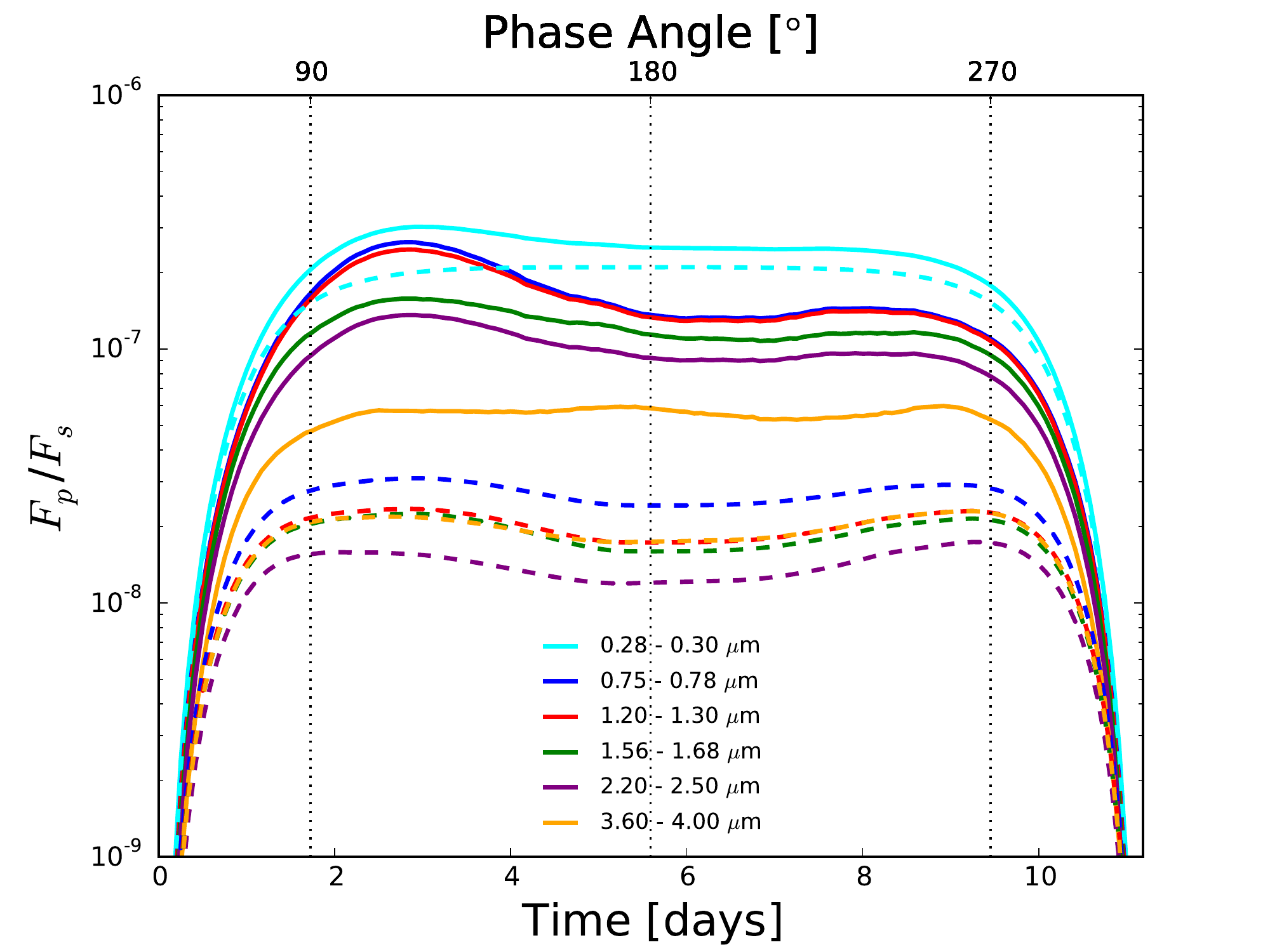}
  \includegraphics[width=0.8\columnwidth]{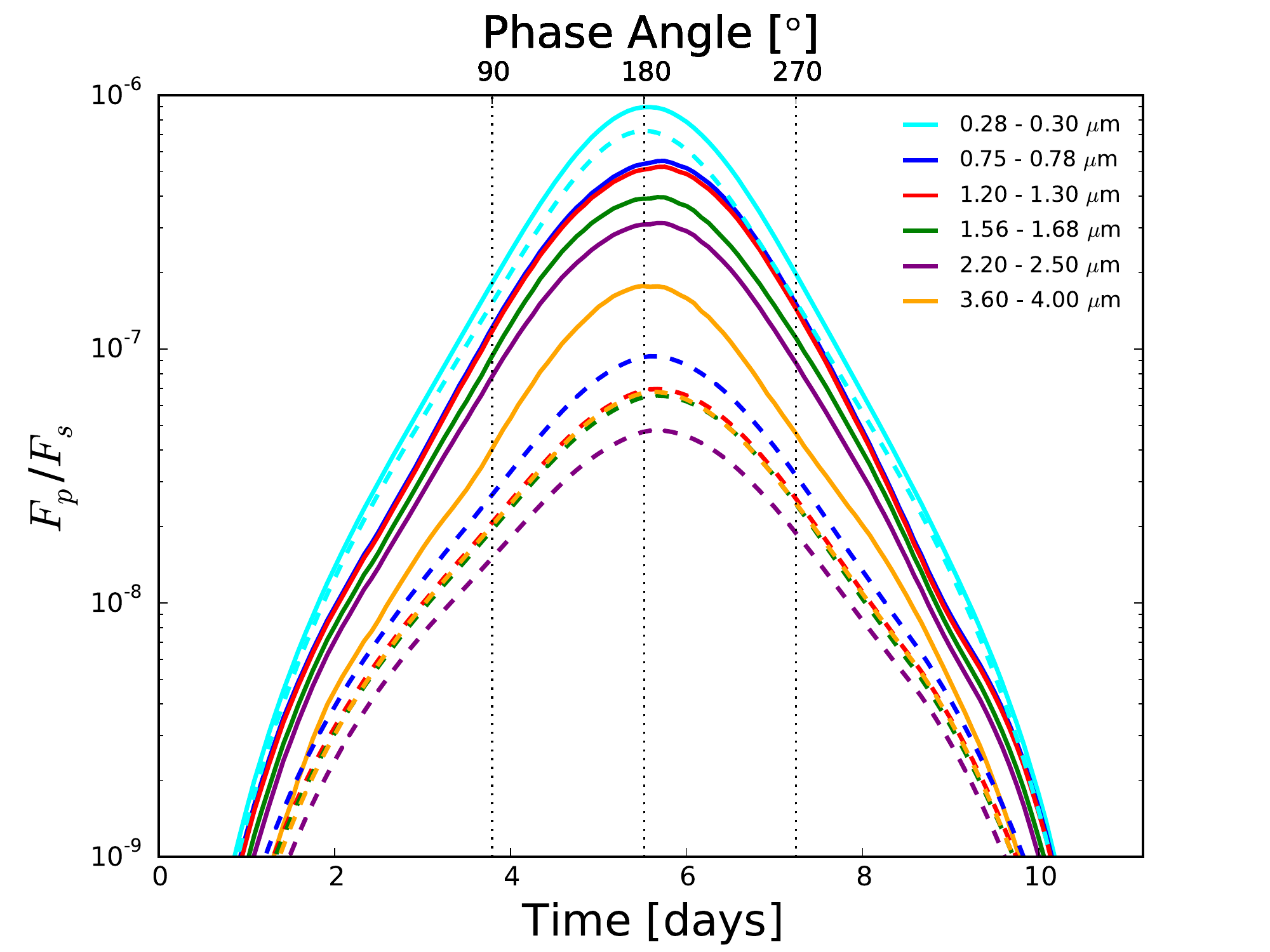}
  \includegraphics[width=0.8\columnwidth]{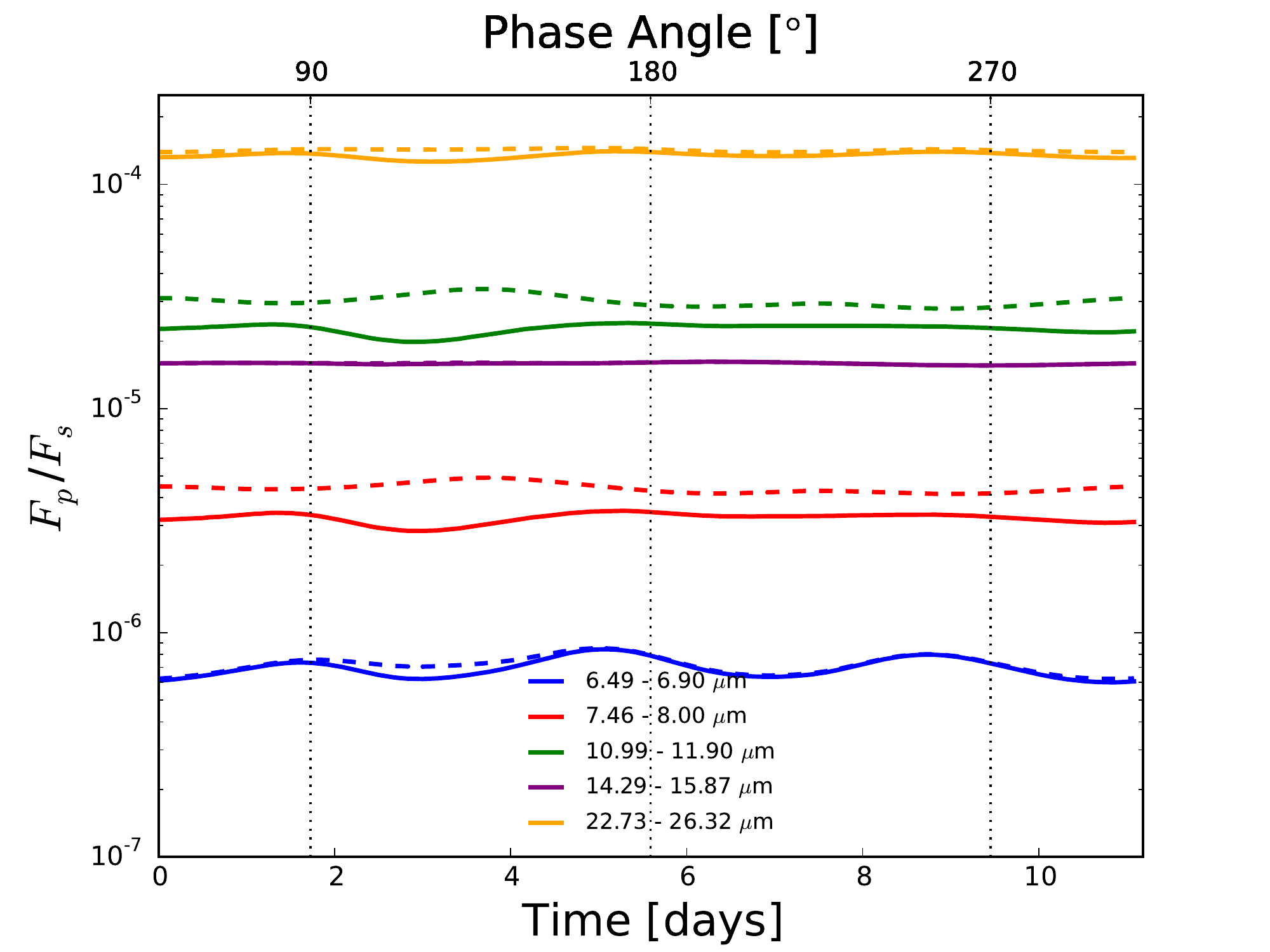}
  \caption{{ Reflectance phase curves for observer longitudes such
      that periastron is at 102$\degree$, 0$\degree$ and 180$\degree$
      phase angle({\it from top to second from bottom}), and the
      emission phase curve for an observer such that periastron is at
      0$\degree$ phase ({\it bottom}) for the 3:2 resonance
      nitrogen-dominated case. Lines are as in
      Fig.~\ref{el_tl_phase}. We only show one orbit, as due to the
      symmetry of the planetary climate, the subsequent orbit produces
      almost identical phase curves. The emission phase curves show
      very little variation with observer longitude, hence we only
      show the view from behind periastron for brevity.}}
  \label{n2_3_2_phase}
\end{figure}

Figure~\ref{el_tl_phase} shows the emission as a function of
orbital phase angle for the Earth-like, tidally-locked
simulation, at three inclinations ($i$=30$\degree$, 60$\degree$ \&
90$\degree$) for the shortwave and at only one inclination
($i$=60$\degree$) for the longwave (due to the invariance of the
longwave phase curve with inclination). { We do not adjust the radius (and therefore total planetary flux) with
inclination using $R_p\propto (M_{min}/\sin i)^{0.27}$ as in \cite{TurLS16}, since to be fully
consistent this would require running all climate simulations with an
adjusted radius. Instead the phase curves represent an identical
planet observed from different angles}. As with Figure~\ref{el_tl_emission} the
``clear-sky'' contribution is also shown as a dashed line.

We generally find very good agreement with the results of
\citet{TurLS16}. Of notable difference in the reflectance phase curves
is the much reduced flux ratio in the 0.28--0.30~$\mu$m region, due to
absorption by ozone which is not included in the model of
\citet{TurLS16}, which means that our predicted flux contrast is two orders of magnitude smaller in this band. In addition, we find the largest flux ratio to be in the
1.20--1.30~$\mu$m band, in contrast to \citet{TurLS16} who find the
0.75--0.78~$\mu$m band to possess the largest contrast (disregarding the previously discussed 0.28--0.30~$\mu$m band). We find the
planetary flux in the 0.75--0.78~$\mu$m is depressed by the oxygen
absorption line at 0.76~$\mu$m, which is not included in the model of
\citet{TurLS16}. Our longwave emission phase curves also show very
similar trends to \citet{TurLS16}. { However, we find that the flux contrast in the bands 7.46--8.00~$\mu$m and 10.9--11.9~$\mu$m is a factor few lower in our model, likely due to the presence of additional absorption by trace gases (CH$_4$ and N$_2$O at 7.46--8.00~$\mu$m) and the cooler surface temperature.}

In Figure~\ref{el_tl_phase} we also show the clear--sky flux, ignoring
the radiative effects of clouds, to highlight the important role that
clouds have on the magnitude of the reflectance
phase curves. The high albedo of clouds results in increases to the
planet-star flux ratio by an order of magnitude. On the other hand,
clouds have a much more subtle direct impact on the longwave emission
spectrum; though of course the temperature of the atmosphere/surface
{ has been influenced by the presence of} clouds, and so they have an important
indirect effect on the longwave emission through the temperature.

Figure~\ref{n2_3_2_emission} shows the reflection and emission spectra
for the nitrogen dominated atmosphere in the eccentric ($e=0.3$), 3:2
resonance orbit. The shortwave spectrum is very similar to the
tidally-locked Earth-like case, though note the lack of ozone
absorption at very short wavelengths; ozone and other trace species
not being included in this model. The longwave spectrum is very
insensitive to { orbital phase} and inclination, due to the
horizontally uniform nature of this atmosphere, as opposed to the
tidally-locked model. The spectrum is generally quite featureless
except for the { absorption feature due to CO$_2$ around 15 $\mu$m.}

Figure~\ref{n2_3_2_phase} shows the emission as a function of
orbital phase for the 3:2 spin-orbit nitrogen dominated
model. { The full repeating pattern should contain two complete orbits, but due to the symmetry of the planetary climate, each orbit produces a very similar phase curve, and therefore we only present a single orbit for clarity.}

For the shortwave reflection, we find broadly similar results to those
of the tidally-locked Earth-like case. { However, in this case the phase curve is now strongly affected by the longitudinal position of the observer and eccentricity of the orbit. It is almost symmetric when viewed from periastron or apoastron, and any deviations from this are due to the atmospheric variability of the planet. For example, the peak flux contrast when viewed from periastron is at $\sim 120\degree$, and is due to reflection from the high, convectively generated cloud above the hot-spot, which was recently heated at periastron and has just appeared into view. There is no corresponding peak at $\sim 240\degree$ because although we are again seeing a hot-spot, this one has not been heated since periastron on the previous orbit, and so the convective cloud has decayed significantly. When viewed from the side, the planetary phase curves display a strong asymmetry. The most striking asymmetry is due to our choice to present the phase curve with a linear time axis (as a distant observer would see) rather than a linear phase angle axis, and is created by the apparent speedup and slowdown of the orbit near periastron and apoastron. However, the phase curve would still be asymmetric if plotted against a linear phase angle axis, due to the variation in stellar radiation received. For example, the peak clear-sky flux is at $\sim 150\degree$, and occurs because the peak radiation is received at periastron ($\sim102\degree$), after which the radiation available for reflection is reducing whilst the visible area of the planet which is illuminated is increasing. The total flux is then offset slightly towards $180\degree$ from this, because there is a delay in the formation of the high, convectively generated clouds above the hot-spot after peak irradiation. The fact that features in the phase curves are dependent on the hydrological cycle, and the formation and evolution of water clouds, hints at an exciting opportunity to constrain this feature given sensitive enough observations.}

The longwave phase curves
show much less variation in the flux with orbital phase compared with the tidally-locked model. This
is despite the fact that this model contains two hot spots due to the
eccentricity ($e=0.3$) of the orbit, { which are visible in the slight increase in contrast at $\sim 150\degree$}. In fact, Figure~\ref{n2_3_2_phase}
shows that the small variations in the flux due to these hot spots are
damped out further by radiative effects of clouds, { which appears as a reduction in the contrast over the hot-spot regions}.

Overall, our simulated observations show many consistencies with those
of \citet{TurLS16}, however, we also find a few important differences. Firstly, our
calculations were performed in a much higher spectral resolution,
allowing us to pick out specific absorption and emission features, in
particular those associated with the gases vital to complex life on Earth,
oxygen, ozone and CO$_2$. { Secondly, we find that the shortwave phase curves show significant asymmetry in the 3:2 resonance model.}

\section{Conclusions}
\label{section:conclusions}

This paper has introduced the use of the Met Office Unified Model for
Earth-like exoplanets. Using this GCM, we have been able to
independently confirm results presented in \cite{TurLS16}, that ProC B
is likely to be habitable for a range of orbital states and
atmospheric compositions. Given the differences between the models,
both in numerics (the fluid equations being solved and how this is
done) and physical parametrizations (the processes included and level
of complexity of the schemes), the level of agreement between the
models is somewhat remarkable. Having this level of agreement from
multiple GCMs is an important factor in the credibility of any results
which are produced with a GCM, especially for cases such as ProC B
where the observational constraints are (currently) very limited. As
with many phenomena on Earth for which observations are limited, an
alternative strategy to constrain the climate simulations could be
more detailed modelling of specific parts of the climate system. For
example, high-resolution convection resolving simulations of the
sub-stellar point of the tidally-locked case would help to constrain
the amount of cloud, precipitation, and export of moisture to the
night side of the planet.

We have additionally shown in this paper that the { range of orbital states for which}
ProC B may { be habitable is} larger than that proposed by \cite{TurLS16}. By
use of a different, weaker, stellar flux, we have shown that the
planet is still easily within { a habitable orbit} despite the known
uncertainty in the luminosity of Proxima Centauri. This is a
consequence of a particularly low sensitivity of planetary
temperatures to changes in stellar flux received for ProC B. The
inclusion of eccentricity to a 3:2 resonant orbital state was also
shown to increase the { habitability}. This result { could} hold true for
any planet near the outer (cold) limit of its habitability zone --
including eccentricity in an orbit { has the potential to} increase the size of
the habitable zone. Two factors combine to produce this
effect. Firstly, the mean stellar flux will always be higher for an
eccentric orbit, and secondly, for orbits in resonant states, a large
increase in the stellar flux is received by fixed regions on the
planet surface, which become permanent `hot-spots'. The circulation on
planets in this configuration is not too dissimilar from the tidally
locked case, with the number of hot-spots being set by the spin-orbit
resonance. A planet in an eccentric orbit with a 2:1 resonance would
have a single hot-spot and be very similar to the tidally-locked
case. Whether eccentric orbits would reduce the habitable zone for
planets near the inner (hot) limit is a matter for further study, and
is likely to depend on other { planetary factors, such as tidal heating discussed by \cite{BolSO17}, and} climate feedbacks such as the cloud
feedbacks discussed by \cite{YanCA13}.

There are obviously several ingredients missing from our analysis. We
have neglected the presence of any land-surface, as we have no
information what this may look like, but in considering the surface to
be water covered we have additionally neglected any transport of heat
by the oceans. \cite{HuY14} have recently investigated tidally-locked
exoplanets with ocean circulation, demonstrating that the ocean acts
to transport heat away from the sub-stellar point. If this were
included in our simulations here, it is likely that the region where
surface temperatures were above freezing would be increased in both
the tidally-locked and 3:2 resonance cases, further reducing the
potential for the 3:2 case to fall into a snowball regime. There is
also the possibility that the location of continents and surface
orography relative to the regions where surface temperatures were
above freezing may significantly affect the planet's climate and
habitability.

We have also considered some more exotic chemical species (oxygen,
ozone, methane etc) in our analysis, but have only specified them as
globally invariant values. The reason for this is their importance in
the evolution of complex life on Earth. The next logical step here is to
couple a chemistry scheme to the atmospheric simulations, that allows
these species to be formed, destroyed and transported by the
atmospheric dynamics, and this would allow a much better estimation of
the likely atmospheric composition.

We have generated, from our simulations, the emissions from the planet
as a function of wavelength in high-resolution, allowing us to
highlight signatures of several key (for complex life on Earth) gaseous
species in the spectrum (oxygen, ozone and carbon dioxide). We have
also generated emissions as a function of { orbital phase angle}
from our simulations and find results largely consistent with
\cite{TurLS16}. { Overall our findings are similar, though our
  results present a higher spectral resolution, and show the
  importance of the observer longitude on the appearance of phase
  curves when the planet has an eccentric orbit.}

\begin{acknowledgements}
  I.B., J.M. and P.E. acknowledge the support of a Met Office Academic
  Partnership secondment. B.D. thanks the University of Exeter for
  support through a Ph.D. studentship. N.J.M. and J.G.'s contributions
  were in part funded by a Leverhulme Trust Research Project Grant,
  and in part by a University of Exeter College of Engineering,
  Mathematics and Physical Sciences studentship. We acknowledge use of
  the MONSooN system, a collaborative facility supplied under the
  Joint Weather and Climate Research Programme, a strategic
  partnership between the Met Office and the Natural Environment
  Research Council. This work also used the University of Exeter
  Supercomputer, a DiRAC Facility jointly funded by STFC, the Large
  Facilities Capital Fund of BIS and the University of Exeter. { We thank an anonymous reviewer for their thorough and insightful comments on the paper, which greatly improved the manuscript.}
\end{acknowledgements}

%
\bibliographystyle{aa} 
\bibliography{bibliography} 
%

\end{document}